\documentclass[twocolumn]{aastex6}

\shorttitle{RR Lyrae stars in the Crater~II dwarf galaxy}
\shortauthors{Joo et al.}

\begin{document}

\title{RR LYRAE VARIABLE STARS IN THE CRATER~II DWARF GALAXY}

\author{
Seok-Joo~Joo\altaffilmark{1}, Jaemann~Kyeong\altaffilmark{1}, Soung-Chul~Yang\altaffilmark{1}, Sang-Il~Han\altaffilmark{1}, Eon-Chang~Sung\altaffilmark{1},
Dongwon~Kim\altaffilmark{2}, Hyunjin~Jeong\altaffilmark{1}, Chang~H.~Ree\altaffilmark{1}, Soo-Chang~Rey\altaffilmark{3}, Helmut~Jerjen\altaffilmark{4},
Hak-Sub~Kim\altaffilmark{1}, Sang-Mok~Cha\altaffilmark{1,5}, and Yongseok~Lee\altaffilmark{1,5}
}

\altaffiltext{1}{Korea Astronomy and Space Science Institute, Daejeon 34055, Korea; sjjoo@kasi.re.kr}
\altaffiltext{2}{Department of Astronomy, University of California, Berkeley, CA 94720-3411, USA}
\altaffiltext{3}{Department of Astronomy and Space Science, Chungnam National University, 99 Daehak-ro, Daejeon 34134, Korea}
\altaffiltext{4}{Research School of Astronomy and Astrophysics, Australian National University, Canberra, ACT 2611, Australia}
\altaffiltext{5}{School of Space Research, Kyung Hee University, Yongin, Kyeonggi 17104, Korea}

\begin{abstract}

We report the detection of RR Lyrae variable stars in Crater~II, a recently discovered large and diffuse satellite dwarf galaxy
of the Milky Way (MW). Based on $B$, $V$ time-series photometry obtained with the Korea Microlensing Telescope Network (KMTNet)
1.6\,-m telescope at CTIO, we identified 83 $ab\,$-type and 13 $c\,$-type pulsators by fitting template light curves.
The detected RR Lyrae stars are centrally concentrated, which ensures that most of them are members of Crater~II.
In terms of the distribution of RRab stars in the period-amplitude diagram, Crater~II is clearly different from ultra-faint dwarf (UFD) galaxies,
but very similar to the two classical MW dwarf spheroidal (dSph) galaxies Draco and Carina with Oosterhoff-intermediate (Oo-int) properties.
Combined with the mean period of $ab\,$-type variables ($\langle P_{ab}\rangle = 0.631\pm 0.004$\,d)
and the $c\,$-type fraction ($\sim$0.14) in Crater~II,
this suggests an Oo-int classification for Crater~II and implies that its nature is more like a dSph rather than a UFD.
We also estimated the mean metallicity, reddening, and distance of Crater~II,
from the photometric and pulsation properties of the RR Lyrae stars.
The stellar population model we have constructed indicates that Crater~II is dominated by an old population,
but is relatively younger than the oldest globular clusters in the MW.
With a lack of high-amplitude short-period RRab stars, Crater~II, like most of the other less massive dSphs,
is probably not a surviving counterpart of the major building blocks of the MW halo.
\end{abstract}


\section{INTRODUCTION} \label{sec:intro}

Over the past dozen years, the number of known satellite galaxies around the Milky Way (MW) has dramatically increased from 12 to $\sim$50
(\citealp[e.g.,][]{wil05a,wil05b,bel06,bel07,bel08,bel09,bel10,gri06,gri09,zuc06a,zuc06b,sak06,irw07,wal07,bec15,drl15,drl16,kim15a,kim15b,kim15,
lae15a,lae15b,mar15,kop15,tor16a,tor16b}; \citealp[see also][updated 2015]{mcc12}; \citealp[][for reviews]{bel13}), thanks to large optical surveys
such as the Sloan Digital Sky Survey (SDSS; \citealt{yor00}), the Dark Energy Survey (DES; \citealt{des16}), the Pan-STARRS1 Surveys \citep{cha16},
and the VLT Survey Telescope (VST) ATLAS survey \citep{sha15}.
Most of these new MW dwarf companions are very faint in terms of both total luminosity ($M_V \gtrsim -8$) and
surface brightness ($\mu_{_V} \gtrsim 28$\,mag\,arcsec$^{-2}$), leading to the term ultra-faint dwarf (UFD) galaxies.
Spectroscopic and/or photometric follow-up studies \citep[e.g.,][]{mun06,mun10,mar07,sim07,sim11,kir08,fre10,nor10,oka12,san12,bro12,bro14,koc14,kim16,con18}
in turn revealed that these least luminous galaxies are also most dark matter dominated (mass-to-light ratio, $M_{\odot}/L_{V,\odot} \gtrsim 100$),
least chemically evolved ([Fe/H] $\lesssim -2.5$), and probably old ($t \gtrsim$~10\,Gyr).
It is now naturally expected from the steep slope of the luminosity function of MW dwarf companions \citep{kop08} that
the universe is dominated by these faintest galaxies in number.

The UFDs are believed to play an important role in resolving the well-known ``missing satellite" problem, the discrepancy between observations
and predictions from the current $\Lambda$CDM hierarchical merging paradigm, for the number and spatial distribution of satellite galaxies
(\citealp{kau93,kly99,moo99}; \citealp[see also][]{bul05,sim07,kop09,bah14}). Hence, not only is a more complete census of MW dwarf satellites
over the entire sky needed, but also detailed studies of their stellar populations are crucial to better understand their true nature,
and thus galaxy formation and evolution via accretion and merger processes \citep[e.g.,][]{wil04,jer10,con18}.

The Crater~II dwarf galaxy is one of the most recently discovered MW satellites in the southern hemisphere, first reported by \citet{tor16a}
using the VST ATLAS survey data. From the total luminosity of $M_{V} \simeq -8.2$, Crater~II can be classified as either one of the brightest UFDs
or one of the faintest classical dwarf spheroidal (dSph) galaxies. Interestingly, given the total luminosity, this galaxy is very large (half-light radius,
$r_{h} \approx 1066$\,pc) and diffuse ($\mu_{_V} \approx 30.6$\,mag\,arcsec$^{-2}$). It is currently the fourth largest MW satellite
--- only the Large Magellanic Cloud (LMC), the Small Magellanic Cloud (SMC), and the Sagittarius (Sgr) dwarf galaxy are larger ---
and one of the lowest surface brightness galaxies known \citep[see their Figure 6]{tor16a}. Based on the color-magnitude diagram (CMD)
and isochrone fitting, \citet{tor16a} estimated Crater~II to have an old age ($\sim$10\,Gyr) and low metallicity ([Fe/H] $\simeq -1.7$).
More recently, from MMT/Hectochelle spectroscopy, \citet{cal17} identified $\sim$62 red giant branch (RGB) stars in this galaxy, and obtained
a mean metallicity of $\langle \rm [Fe/H] \rangle \simeq -1.98$ with a dispersion of $\sigma_{\rm[Fe/H]} \approx 0.22$.
They also showed that it has an extremely low line-of-sight velocity dispersion ($\sigma_{v,\rm los} \approx 2.7$\,km\,s$^{-1}$, \citealp[see also][]{mcg16})
and a dynamical mass of $\sim$$4.4 \times 10^6 M_{\odot}$,
suggesting a mass-to-light ratio of $\sim$53 $M_{\odot}/L_{V,\odot}$ within the half-light radius.
Located relatively distant from the Sun ($\gtrsim$100\,kpc) and widely distributed on the sky ($r_{h} \simeq 31\,'$, \citealt{tor16a}),
there is still a lack of extensive studies on the stellar population of Crater~II.

As radially pulsating low-mass horizontal branch (HB) stars in the phase of core helium burning,
RR Lyrae variables are good tracers of old ($\gtrsim$10\,Gyr) and metal-poor populations (i.e., population~II).
Their photometric and pulsation properties including mean magnitude, period, amplitude, and light curve, are commonly used to derive
metallicity, interstellar reddening, and distance of the system.
They can also provide information on the stellar structure, formation, and evolution of their host dwarf galaxies,
particularly when compared with those stars in the Galactic halo or globular clusters (GCs) \citep[e.g.,][]{smi04,cat09,cle10,pie15}.
Since 2006, about 13 UFDs (with $M_V > -7$) have been reported to contain RR Lyrae variables \citep[see their Table 4 for a recent compilation]{viv16},
including Bo\"otes~I (holding the current record with 15 RR Lyrae stars) and Segue~I, one of the faintest UFDs known \citep{sim11}.
It is interesting to note that all dwarf galaxies searched for variable stars so far have at least one RR Lyrae star
(\citealp{viv16}; \citealp[see also][]{gar13,boe13,ses14,med17}) regardless of the total luminosity.

In this paper, we investigate the RR Lyrae population in the Crater~II dwarf galaxy based on time-series observations using
the Korea Microlensing Telescope Network (KMTNet) 1.6\,-m telescope located at the Cerro-Tololo Inter-American Observatory (CTIO).
This study is part of our ongoing southern hemisphere survey for RR Lyrae stars in UFDs using the KMTNet-CTIO.
Section\,2 presents the time-series observations and data reduction process.
Section\,3 describes our detection of the RR Lyrae stars
by applying the template light curve fitting routine, RRFIT, developed by \citet{yan12},
and the characterization of the stars including their light curves, spatial distribution and metallicity estimates.
We discuss the results and draw our conclusions in section\,4.

\section{OBSERVATIONS AND DATA REDUCTION} \label{sec:obs}

Time-series $B$, $V$ observations of the Crater~II dwarf galaxy were carried out using the KMTNet-CTIO 1.6-m telescope
during 2016 February 1--4, 6--10, and 2017 January 29--30 (UT). The telescope is equipped with the mosaic CCD camera of 18k $\times$ 18k pixels,
providing a wide field of view of $2 \degr \times 2 \degr$ and a pixel scale of 0.40$\,''$ \citep{slkim16}.
Our observations covered a total area of $\sim$$3 \degr \times 3 \degr$ with five largely overlapping fields, which is
more than twice the half-light radius of the galaxy ($r_{h} \simeq 31\,'$, \citealt{tor16a}).
With an exposure time of 120\,s per image, we obtained in total 143 and 145 frames in the $B$- and $V$-bands, respectively,
which correspond to a total exposure time of $\sim$4.8\,hr in each passband.

All images were preprocessed by the KMTNet pipeline which includes basic corrections such as cross-talk, bias subtraction, and flat fielding \citep{slkim16}.
We performed point-spread function (PSF) photometry using DAOPHOT II/ALLSTAR and ALLFRAME \citep{ste87,ste94},
similarly to the previous works on Galactic GCs \citep{han09b,han15}.
Photometric calibration was achieved using the AAVSO Photometric All-Sky Survey (APASS) database (\url{http://www.aavso.org/apass}),
and astrometric calibration was performed with SCAMP \citep{ber06} applying the third US Naval Observatory (USNO) CCD Astrograph Catalog, UCAC3 \citep{zac10}.

\begin{figure*}
\epsscale{1.0}
\plotone{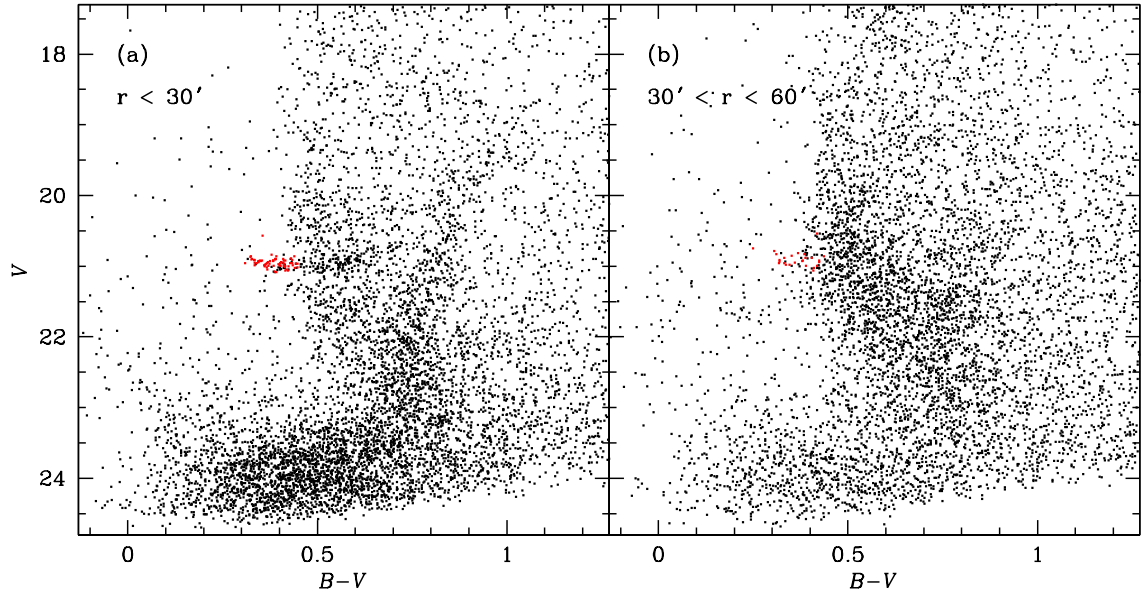}
\caption{
CMDs of stars within and outside the half-light radius ($r_{h} \simeq 31'$, \citealt{tor16a}) of Crater~II,
obtained from the combined images of our time-series data.
The CMD features of Crater~II such as the red HB, RGB, SGB, and MSTO, are clearly visible in panel (a).
The 96 RR Lyrae stars we detected are highlighted as red points. See also Figure~\ref{fig8}(b) for their spatial distribution.
\label{fig1}}
\end{figure*}

The panels in Figure~\ref{fig1} show the CMDs of stars within and outside the half-light radius ($r_{h} \simeq 31'$) of Crater~II,
produced by performing PSF photometry on the combined images of all our time-series data with SWarp \citep{ber02}.
The CMDs were cleaned of poorly measured stars and non-stellar objects using the photometric errors and the ALLFRAME fitting parameters, CHI and SHARP,
as functions of $B$ and $V$ magnitudes \citep[see][]{kim13,lim16}.
In panel~(a), for stars within the half-light radius,
we can clearly see the CMD features of Crater~II, such as the RGB, red HB, sub-giant branch (SGB), and even main-sequence turn-off (MSTO).
Panel~(b), for stars outside the half-light radius, however, shows only weak features of Crater~II superimposed on the background contamination,
indicating that most member stars are within the half-light radius.
Figure~\ref{fig1} generally confirms the CMD properties of Crater~II presented by \citet[see their Figure 1]{tor16a},
but our observations are deep enough to reach the MSTO near $V \simeq 24$,
which is important for age dating of stellar populations (see Section~\ref{sec:discussion}).
The red dots represent RR Lyrae stars identified in this study, where their colors and magnitudes are mean values from our light curve analysis (see Section~\ref{sec:rrl}).

\section{RR Lyrae variable stars} \label{sec:rrl}

\subsection{Detection and Characterization} \label{subsec:detect}

To identify and characterize RR Lyrae stars in the field of Crater~II, we applied the template light curve fitting routine,
RRFIT, developed by \citet{yan12} on the basis of template fitting methods by \citet{lay98} and \citet{man08}.
Following the technique of \citet{yan10,yan14} and \citet{yan12},
we first selected stars at the HB luminosity level, i.e., $20.2 < V < 21.7$,
and examined their variability using the reduced chi-square, $\chi^2_\nu$, defined as,
\begin{equation}\label{eq1}
    \chi^2_\nu = \frac{1}{N_B+N_V} \times \left[ \sum_{i=1}^{N_B}\frac{(B_i-\overline{B})^2}{\sigma_{Bi}^2} + \sum_{i=1}^{N_V}\frac{(V_i-\overline{V})^2}{\sigma_{Vi}^2} \right] ,
\end{equation}
where, for each star in the $B$- and $V$-bands, $N_B$ and $N_V$ are the numbers of the observed frames, $B_i$ and $V_i$ are the apparent magnitudes in the $i_{th}$ images
with uncertainties $\sigma_{Bi}$ and $\sigma_{Vi}$, and $\overline{B}$ and $\overline{V}$ are the mean magnitudes for the $N_B$ and $N_V$ frames.
In the calculation of the $\chi^2_\nu$ value, data points further away than 3\,$\sigma$ from the mean magnitude were excluded.
Variable stars are expected to have larger $\chi^2_\nu$ values than typical non-variable stars that have values around unity ($\chi^2_\nu \approx 1$) in an ideal situation.
We considered the stars with $\chi^2_\nu > 3.0$ as potential variable candidates, to be about 30\% of those selected earlier to be at the HB level.

We have then analyzed the time-series data of these variable candidates using RRFIT,
where we rejected data points with magnitude error greater than 0.1.
Based on template light curves of RR Lyrae stars, RRFIT determines the best-fit light curve and parameter values
such as period, amplitude, mean magnitude, and epoch of maximum light.
As in the previous works by \citet{yan10,yan14} and \citet{yan12}, we used a library of 25 light curve templates,
which are the 6 $ab\,$-type and 2 $c\,$-type templates from \citet{lay00} and the 17 $ab\,$-type templates from \citet{kov07}.
By visually inspecting the output light curves from RRFIT, we finally identified 96 RR Lyrae stars,
including 83 fundamental mode ($ab\,$-type) and 13 first overtone ($c\,$-type) pulsators.\footnotemark[6]
Figures~\ref{fig2} to \ref{fig6} present the light curves and the best-fitting templates for all the detected RR Lyrae stars.
Table~\ref{tbl1} lists their pulsation properties,
including coordinate, type, period, epoch of maximum light, intensity-weighted mean magnitude ($\langle B\,\rangle$ and $\langle V \rangle$),
the number of observations ($N_B$ and $N_V$), and amplitude ($A_B$ and $A_V$).
The mean magnitudes are obtained by averaging the intensity of the best-fit templates over 0.02 phase interval.
The quantities $N_B$ and $N_V$ are the actual numbers of data points used in RRFIT, and some stars, for example those lying
in the gaps between the CCDs, have fewer data points compared to stars detected in all observations.

\footnotetext[6]{The faintest variable star detected in the field, V97, is not counted here since it is probably not an RR Lyrae star (see Section~\ref{subsec:cmd}).
}

\subsection{Synthetic Light Curve Simulation} \label{subsec:syntetic}

Time-series analysis may suffer from aliasing (i.e., spurious periods), which is mainly caused by the limited number of observations,
poor phase coverage, and/or photometric uncertainty.
Hence, we need to scrutinize at what level the aliases affected the pulsation periods before using them to derive
metallicity, reddening, and distance of Crater~II.
For this purpose we performed synthetic light curve simulations and statistical tests following the prescription of \citet{yan14}.
Based on the light curve templates of \citet{lay00} for RR Lyrae stars, we first generated 3,000 $ab\,$-type and 2,000 $c\,$-type artificial light curves
by applying our observational constraints (such as number of epochs, cadence, observing baseline,
and photometric errors) which were extracted from a few of the good RR Lyrae candidates.
The (input) periods and amplitudes were randomly assigned to each artificial RR Lyrae star
within the appropriate ranges of $ab\,$- and $c\,$-type variables, respectively.
We then ran RRFIT on these synthetic time-series data in the same way as for the observed ones,
so that we can directly compare the assigned (i.e., input) and calculated (i.e., output) parameters.

The top and bottom panels in Figure~\ref{fig7} present the difference between input and output periods
as a function of the input periods, for RRab and RRc stars, respectively.
Our simulations show that $c\,$-type variables are more affected by aliasing than $ab\,$-type stars
in the sense that the output periods are generally longer than the input periods.
While the input periods of $\sim$91\% of the synthetic RRab stars were recovered within $\pm$0.05\,d,
only those of  $\sim$27\% of the artificial RRc stars were recovered within the same period range.
Consequently, we have less confidence in the pulsation parameters calculated for the RRc stars in this study,
and decided to use them only for the the mean magnitude and the $c\,$-type fraction (see Section~\ref{subsec:cmd}).

We also estimated the systematic error in the mean period introduced by our period searching method based on statistical analysis.
By employing the synthetic RR Lyrae (i.e., 3,000 RRab and 2,000 RRc) stars as a parent population,
we randomly selected a subsample with the same number of stars as detected in our observation (i.e., 83 RRab and 13 RRc stars).
Since each artificial RR Lyrae has a $\Delta P$ (input - output) value associated, we can calculate an average $\Delta P$ value for the subsample
(i.e., $\langle \Delta P \,\rangle_i$ ). To increase the statistical significance, we repeated this random sampling 10,000 times
and produced the $\langle \Delta P \,\rangle$ distribution for the 10,000 subsamples.
We consider the 1\,$\sigma$ range of the $\langle \Delta P \,\rangle$ distribution from the best-fit Gaussian as a good estimate for the systematic error
in the mean period of the RR Lyrae stars derived by our template light curve analysis. We find a systematic error,
$\sigma_{\langle P_{ab} \rangle} = \pm$0.0018\,d for the mean period of the RRab stars.

\subsection{CMD and spatial distribution} \label{subsec:cmd}

The panels in Figure~\ref{fig8} present the CMD, spatial distribution, and color-period diagram of the RR Lyrae stars we have detected in Crater~II.
The two grey vertical lines in the upper panels denote the empirical instability strip,
roughly estimated by averaging the blue and red boundaries of 9 Galactic and LMC clusters from \citet[][see his Table~7]{wal98}.
The boundaries were reddened by assuming an interstellar extinction value of $E(B-V) = 0.05$\,mag (see Section~\ref{subsec:reddening}).
It is clear from panel~(a) and Figure~\ref{fig1} that most RR Lyrae stars are within the appropriate color range of the instability strip
at the level of the HB.
Panel~(a) also shows that the RR Lyrae stars gather on the red side of the instability strip in their color distribution.
This further indicates that Crater~II has a red HB morphology
with virtually no blue HB stars, which leads the HB morphology index of \citet{lee94}, $(B-R)/(B+V+R)$, to be roughly $\lesssim$$-0.5$,
where $B$, $V$, and $R$ are the numbers of blue HB, variable (RR Lyrae) and red HB stars, respectively.
This red HB morphology naturally explains the relatively small number of $c\,$-type variables compared to that of $ab\,$-type stars,
yielding the $c\,$-type fraction to all RR Lyrae variables, $N(c)/N(ab+c) \simeq 0.14$.

Panel~(b) illustrates that the RR Lyrae stars are clustered around the galaxy's center estimated by \citet{tor16a}.
The two histograms on the top and right sides of the panel demonstrate that they are centrally well concentrated.
About 70\% (67 out of 96) of RR Lyrae stars are within $30\,'$ (approximately the half-light radius),
and about 97\% (93 out of 96) of them are within $60\,'$ from the center.
The CMD and spatial distribution of the RR Lyrae variables in panels~(a) and (b) thus strongly suggest that most of them belong to Crater~II.
The large heliocentric distance of the galaxy ($\sim$112 kpc, see Section~\ref{subsec:reddening}) further ensures this interpretation,
because the Galactic halo RR Lyrae stars are thought to be rare at distances greater than $\sim$80 kpc \citep{viv16,zin14,wat09}.
The three stars (cyan open circles) located more than $60\,'$ ($\gtrsim 2\,r_{h}$) from the center, on the other hand, are also
among the brightest (V37, V79) or the reddest (V96) variables in the CMD. These three outliers might be field stars,
and are denoted as ``field?" in Table~\ref{tbl1}.

In panel~(c), we plot the fundamental periods (P$_f$) of the RR Lyrae stars as a function of $B-V$ color,
where the periods of RRc stars were fundamentalized assuming the period ratio between the $c\,$-type and $ab\,$-type stars,
$P_c / P_{ab}$ = 0.745 \citep{cle01,nem85}. While most RR Lyrae stars follow a tight correlation between the fundamental
period and color, some stars (green symbols) have considerably longer or shorter periods at a given color compared to the other stars.
The two $ab\,$-type stars with the longest periods, V1 and V26 (green filled circles), are also the most luminous
($\sim$0.4\,mag brighter than the bulk of the variables) as shown in panel~(a).
Given their periods and luminosity, these two stars appear to be either field RR Lyrae stars or
highly evolved RR Lyrae stars from the zero-age HB \citep[see, e.g.,][]{lee90}.
Otherwise, they might be faint anomalous Cepheids (ACs) \citep{pri02,sos08}.
The four $ab\,$-type stars with long periods (green open circles) can be seen as evolved RR Lyrae stars,
since they are also relatively bright in the CMD.
The four $c\,$-type stars with short periods (green open triangles), however, are less certain.
They are too red to be $c\,$-type stars (V69, V93, and V94) or have somewhat short periods (V80).
These four $c\,$-type stars might be field stars or influenced by other factors such as photometric errors or aliasing.
The 10 outliers in panel~(c) are represented in Table~\ref{tbl1}
as either ``field? highly evolved? AC?", ``evolved?", or ``less certain" according to their properties.

For all the 96 RR Lyrae stars, we obtain the mean apparent magnitude, $\langle V_{\rm RR} \rangle = 20.95 \pm 0.01$\,mag, by fitting a Gaussian profile
at the magnitude distribution, where the uncertainty is the standard error of the mean.
Note that, even if the 13 outliers (the three and 10 outliers in panels~(b) and (c), respectively) are excluded,
$\langle V_{\rm RR} \rangle$ does not change.

The faintest variable, V97 (black circle), in panel~(a) is also distinct from the other stars.
It is not only separated from the rest of the RR Lyrae stars in the CMD by its faint magnitude ($\langle V\rangle = 21.334$\,mag),
but also has the shortest period (P~$\simeq 0.235$\,d), even though classified as $ab\,$-type from the shape of the light curve
(see Figure~\ref{fig6} and Table~\ref{tbl1}). With this short period and pulsation mode, V97 might be, not an RR Lyrae star,
but a dwarf Cepheid (DC) \citep[e.g.,][]{mat98,bre00,mcn11} that probably belongs to the MW; it was excluded from our analysis.

\subsection{Period-Amplitude Diagrams} \label{subsec:pad}

It is widely recognized that in contrast to the Galactic GCs, which present the well-known Oosterhoff dichotomy in the average period of RRab stars
and their location in the period-amplitude (Bailey) diagram, the classical dSph galaxies have preferentially Oosterhoff-intermediate (Oo-int) properties
\citep[e.g.,][for reviews]{cat09,cle10,smi11}.
Most RR Lyrae stars in the UFDs studied so far, on the other hand, are classified as Oo-int or Oosterhoff group II (Oo~II) \citep[e.g.,][and references therein]{cle14,viv16}.
Hence, a reliable Oosterhoff classification of Crater~II based on the mean period of RRab stars and the Bailey diagram would help
to understand the properties of its stellar population.

Figure~\ref{fig9} shows the period distribution and the Bailey diagram of the RR Lyrae stars in Crater~II.
We see in panel~(a) that the RR Lyrae stars with the two different pulsation modes are well separated by their periods.
The solid and dotted lines in panel~(b) are the loci of Oosterhoff group I (Oo~I) and Oo~II clusters, respectively,
according to the relation given by \citet[][see also \citealt{cac05}]{zor10}.
The average periods of 83 $ab\,$-type and 13 $c\,$-type stars are $\langle P_{ab}\rangle = 0.631 \pm 0.004$\,d and
$\langle P_c\rangle = 0.411 \pm 0.009$\,d, respectively, where the uncertainties are the standard errors of the means.
If the 13 outliers (nine RRab and four RRc variables) are excluded, the mean periods of 74 $ab\,$-type and nine $c\,$-type stars
slightly change to $\langle P_{ab}\rangle = 0.621 \pm 0.003$\,d and $\langle P_c\rangle = 0.423 \pm 0.004$\,d, respectively.
Note that these $\langle P_{ab}\rangle$ values are in the range of Oo II group clusters, over the Oosterhoff gap
(0.58\,d $\leq \langle P_{ab}\rangle \leq$ 0.62\,d, \citealt{cat09}).
In the period-amplitude diagram, however, the Crater~II RRab stars are located near the Oo~I line,
suggesting that Crater~II may be classified as Oo~I or Oo-int.

In Figure~\ref{fig10}, we compare the RR Lyrae stars in Crater~II with those in the Galactic halo field (data from \citealt{zin14}),
the other 14 UFDs (top panel), and the two classical dSphs, Draco and Carina (bottom panel), on the period-amplitude diagrams.
Note that panel~(a) is very similar to Figure~10 of \citet{viv16}, while the two brightest UFDs with $M_V < -7$,
Canes~Venatici~I (CVn~I) and Leo~T, are included here too.\footnotemark[7]
Data for RR Lyrae variables in the UFDs are taken from \citet[][CVn~I]{kue08}, \citet[][Canes~Venatici~II, CVn~II]{gre08},
\citet[][Coma~Berenices, ComBer]{mus09}, \citet[][Bo\"otes~I, see also \citealt{dal06}]{sie06}, \citet[][Hydra~II]{viv16},
\citet[][Ursa~Major~I, UMa~I]{gar13}, \citet[][Leo~IV]{mor09}, \citet[][Leo~T]{cle12}, and \citet[][Segue~II]{boe13}.
For Bo\"otes~II, Bo\"otes~III \citep{ses14}, and Ursa~Major~II (UMa~II) \citep{dal12}, we used the revised data by \citet{viv16}.
For Bo\"otes~I, the $A_B$ values were converted to $A_V$ using equations derived by \citet{dor99} in the same way as \citet{viv16}.
In the case of Hercules, we added three (1 RRab and 2 RRc) stars recently identified by \citet{gar18} to the data of \citet{mus12},
where the SDSS $g$-band amplitudes, $A_g$, of these three variables were transformed to $A_V$, using the equation,
$V = g - 0.59\,(g-r) - 0.01$, by \citet[see their Table~1]{jes05}.
Similarly, without color information, the $A_g$ values of three RRab stars in Leo~V \citep{med17}, were converted to $A_V$, by assuming $(g-r) = 0.15$
as a rough estimate for the mean color of the RRab stars \citep{an08}.

As presented in panel~(a), the RRab stars in the UFDs (blue symbols), except for Crater~II and CVn~I, are broadly scattered in the Oo-int and Oo~II regions,
while most of those ($\sim$73$\%$) in the Galactic halo field belong to Oo~I \citep{zin14} with some hints of the Oosterhoff dichotomy.
Crater~II and CVn~I, on the other hand, are clearly distinct from the other UFDs, but rather similar to the two classical dSphs with Oo-int properties,
Draco (data from \citealt{kin08}) and Carina (data from \citealt{cop15}), shown in panel~(b).
It is notable that, not only the distribution of RRab stars in the Bailey diagram, but also the mean period of RRab stars ($\langle P_{ab}\rangle$ = 0.631\,d),
and the $c\,$-type fraction ($N(c)/N(ab+c) \simeq 0.14$) of Crater~II are comparable to those of CVn~I and the two classical dSphs, respectively,
i.e., $\langle P_{ab}\rangle$ = 0.60\,d and $N(c)/N(ab+c) = 0.22$ for CVn~I, $\langle P_{ab}\rangle$ = 0.615\,d and $N(c+d)/N(ab+c+d) = 0.21$ for Draco,
and $\langle P_{ab}\rangle$ = 0.634\,d and $N(c+d)/N(ab+c+d)$ = 0.23 for Carina, where we included Blazhko RRab stars and double-mode ($d\,$-type) pulsators
\citep[see also][for the several other dSphs]{ste14,bak15}.
These similarities undoubtedly suggest an Oo-int classification for Crater~II, even though it has a somewhat long $\langle P_{ab}\rangle$,
which corresponds to the Oo~II group.
This also leads us to conclude that, based on the RR Lyrae properties, Crater~II is more like a classical dSph rather than a UFD.

\footnotetext[7]{CVn~I is often considered to be a classical dSph rather than a UFD, because of its classical-dSph-like properties
such as the total magnitude, broad RGB, Oo-int classification, half-light radius, and distribution of alpha-elements \citep[e.g.,][]{sim07,kue08,mar08,san12,var13}.
Leo~T is probably not bound to the MW, located at a large heliocentric distance of $\sim$409 kpc \citep{irw07,dej08,cle12,mcc12}.
Segue~1 is not included here. While one or two RR Lyrae stars are detected by \citet{sim11}, their periods and amplitudes are not accurately measured \citep[see also][]{viv16}.
\\\\
}

\subsection{Metallicity} \label{subsec:metallicity}

RR Lyrae stars can further be used to estimate metallicity, reddening, and distance of a stellar system independently of other methods, since their
pulsation properties (including period and amplitude) are correlated with the stellar evolution parameters such as mass, metallicity, temperature, and luminosity
\citep[see, e.g.,][]{van71,san93,san06,dic04,bon07,jef11}. To obtain metallicities for the individual RRab stars of Crater~II,
we used the empirical period-amplitude-metallicity relation derived by \citet{alc00},
\begin{equation}\label{eq2}
     \rm{[Fe/H]} = -8.85~(log\, \it P_{ab} + \rm 0.15\,\it A_V ) - \rm 2.60 \,,
\end{equation}
where [Fe/H] is on the \citet{zin84} scale with an accuracy of $\sigma_{\rm [Fe/H]}$ = 0.31\,dex.
The upper panel of Figure~\ref{fig11} shows the metallicity distribution of the RRab stars calculated from this equation and a Gaussian fit to the histogram,
which gives the mean metallicity, $\langle \rm[Fe/H]\rangle$ = $-1.65 \,\pm\, 0.15 \,(0.02)$.
The uncertainty is the standard deviation and the number in parentheses presents the standard error of the mean.

This metallicity is in good agreement with the estimation from the isochrone fitting ([Fe/H] = $-1.7 \,\pm\, 0.1$) by \citet{tor16a},
but somewhat more metal-rich than the value from the spectroscopic measurement ($\langle$[Fe/H]$\rangle$ = $-1.98 \,\pm\, 0.1$) by \citet{cal17}.
Note however that, as \citet{cal17} already stated, their estimation might be systematically metal-poor probably due to the zero-point uncertainty of metallicity.
While the relation of \citet{alc00} which we adopt here is more reliable than the other methods based only on the mean period of RRab stars
\citep[e.g.,][]{san06,sar06},
it should be noted that the derived [Fe/H] value has also intrinsically a large dispersion and uncertainty.
For example, it can be affected by the luminosity (evolution) effect on the period of RRab stars \citep{yan10} and/or the selection of calibration cluster \citep{jef11,bon07}.

\subsection{Reddening and Distance} \label{subsec:reddening}

\citet{stu66} has shown that $ab\,$-type RR Lyrae stars have nearly identical intrinsic colors, $(B-V)_0$, in the phase interval from 0.5 to 0.8
(i.e., at minimum light), only weakly correlated with period and metallic-line blanketing.
By combining the Sturch's formula with the calibration between metallicity and line blanketing effect from \citet{but75},
\citet{wal90} presented the following relation for the interstellar extinction,
\begin{equation}\label{eq3}
     E(B-V) = (B-V)_{\rm min} - 0.24 \,P_{ab} - 0.056 \, {\rm [Fe/H]} - 0.336 \,,
\end{equation}
where $(B-V)_{\rm min}$ is the average color at the minimum light (phase between 0.5 and 0.8) and the metallicity scale is that of \citet{zin84}.
Adopting the metallicities from the previous subsection, we obtain reddening values of the individual RRab stars.
The distribution of $E(B-V)$ values and a Gaussian fit at the histogram are plotted (black solid lines) in panel~(b) of Figure~\ref{fig11},
which yields $\langle E(B-V) \rangle$ = $0.09 \pm\, 0.02 \,(0.003)$,
where the uncertainty is the standard deviation and the value in parentheses is the standard error of the mean.
As already noted by \citet[][see also \citealt{pie02,cle03,mus12}]{wal98}, the Sturch's method tends to overestimate the extinction by $0.02 - 0.03$\,mag.
If this effect is taken into account, $\langle E(B-V) \rangle$ would be reduced to $\sim$0.06.
This corrected value is, however, still larger than the estimation based on the \citet{sch98} maps, from which we obtained $\langle E(B-V) \rangle$ = 0.034 for $\rm r < 30\,'$ and
$\langle E(B-V) \rangle$ = 0.037 for $\rm r < 60\,'$ areas around the center of Crater~II (see solid and dotted grey circles in Figure~\ref{fig8}b).

Another independent estimation for the interstellar reddening can be made using the $B$-band amplitude (A$_B$), metallicity, and period of RRab stars, namely,
the amplitude-color-metallicity (ACZ) and/or period-amplitude-color-metallicity (PACZ) relations derived by \citet{pie02},
\begin{equation} \label{eq4}
     (B-V)_0 = 0.448 - 0.078\,A_B + 0.012 \,\rm[Fe/H] \,,
\end{equation}
\begin{equation} \label{eq5}
     (B-V)_0 = 0.507 - 0.052\,A_B + 0.223 \,{\rm log} \,P_{ab} + 0.036 \,\rm[Fe/H] \,.
\end{equation}
The $E(B-V)$ values are then calculated by subtracting these intrinsic colors, $(B-V)_0$, from the magnitude-mean colors, $(B-V)_m$.
The histograms of reddening values obtained from these two relations and Gaussian fits for them are also plotted in panel~(b) in Figure~\ref{fig11}
(red and blue colors, respectively). The peaks of the distributions give $\langle E(B-V) \rangle$ = $0.05 \pm\, 0.02 \,(0.003)$
for both ACZ and PACZ relations, where the uncertainty is the standard deviation with the standard error of the mean in parentheses.
This is roughly in the middle of the values obtained from Sturch's method (when shifted by $-$0.03 mag) and the \citet{sch98} maps,
and therefore we adopt this as our final estimate for the reddening (see Table~\ref{tbl2}).

Once metallicity and reddening are determined, we can estimate the distance to the galaxy from the luminosity-metallicity relation of RR Lyrae stars.
There have been many attempts to establish a universal $M_V \rm(RR) - [Fe/H]$ relation, but some discrepancies still remain between the methods
\citep[see, e.g.,][for reviews]{cha99,cac03,cat09}. We adopt here the slope of the relation, $\Delta M_V \rm(RR) / \Delta [Fe/H] = 0.23~(\pm 0.04)$,
and the zero point, $M_V = 0.56~(\pm 0.12)$ at $\rm [Fe/H] = -1.6$, from \citet[see his equation~8]{cha99}. The relation is then,
\begin{equation}\label{eq6}
     M_V(\rm RR) = 0.23 \,\rm [Fe/H] + 0.93,
\end{equation}
and yields the absolute magnitude $M_V(\rm RR) = 0.55 \pm 0.07$, by applying the mean metallicity of RRab stars derived above,
$\langle \rm [Fe/H] \rangle = -1.65 \,\pm\, 0.15 \,(0.02)$.
The uncertainty in $M_V$ was propagated from the intrinsic deviation of the \citet{alc00} relation.

Combining this with the reddening estimated above and the mean apparent $V$-magnitude, $\langle V_{\rm RR} \rangle = 20.95 \pm 0.01$,
we finally obtain a distance modulus of $(m-M)_0 = 20.25 \pm 0.10$ and
a distance of $\rm d_{\odot} = 112 \pm 5$\,kpc to Crater~II (see Table~\ref{tbl2}). The uncertainty of $(m-M)_0$ is the quadratic sum of the errors
in the $M_V(\rm RR)$ and the $V$-band extinction, $\sim$$3.1 \,E(B-V)$.
Note that this distance estimate agrees with the suggestion by \citet[][ $\rm d_{\odot} = 117.5 \pm 1.1$\,kpc]{tor16a} within the uncertainty.

\section{DISCUSSION} \label{sec:discussion}

Using the 1.6m wide-field KMTNet-CTIO telescope, we performed time-series $B$, $V$ photometry of Crater~II,
one of the largest and lowest surface brightness dwarf satellites of the MW.
We have detected and characterized 96 RR Lyrae stars (83 RRab and 13 RRc) with the template light curve fitting routine, RRFIT,
by \citet{yan12} and \citet{yan10,yan14}.
The mean period of RRab stars, $\langle P_{ab}\rangle = 0.631\pm 0.004$\,d, is somewhat longer than the Oosterhoff gap,
but the $c\,$-type fraction is low ($\sim$0.14) and the location of $ab\,$-type variables in the period-amplitude diagram is close to the locus of Oo~I group clusters.
In terms of these Oosterhoff properties, Crater~II is very similar to CVn~I and the two classical dSph, Draco and Carina.
This not only suggests an Oo-int classification for this galaxy, but also leads us to conclude that it can be categorized
as a classical dSph rather than a UFD.

These similarities in RR Lyrae stars between Crater~II and the classical dSphs indicate that the RR Lyrae properties are independent of the surface brightness of the galaxy.
Instead, given the fact that Crater~II and CVn~I, with $M_V \approx -8.2$ and $-8.6$, respectively, are much more luminous than typical UFDs
and are comparable to the faint classical dSphs, like Draco and Carina with $M_V \approx -8.8$ and $-9.3$\footnotemark[8], respectively \citep{mar08,san12,mcc12,tor16a},
the RR Lyrae properties appear to be more related to the total luminosity of the galaxy.
This interpretation also agrees well with the luminosity-metallicity (or mass-metallicity) relation of (dwarf) galaxies
(\citealp[e.g.,][]{gre03,kir08,wil12,con18}; \citealp[see also][for a recent compilation of nearby dwarfs]{yan14}),
in the sense that stellar population properties are correlated with the luminosity (stellar mass) of the system.

\footnotetext[8]{Unlike the other dwarfs considered here, Carina has a sizeable amount of intermediate-age population with a red HB
that cannot produce RR Lyrae stars \citep{mon03}. If we assume that roughly half of the population in Carina are old enough to host RR Lyrae stars,
then the total magnitude of the old population only would be $M_V \simeq -8.5$.
\\
}

We have then constructed stellar population models, to investigate the star formation history of Crater~II,
following the techniques outlined by \citet{lee90,lee94} and \citet{joo13}.
We used Yonsei-Yale (Y$^2$) isochrones and HB evolutionary tracks \citep{yi08,han09a}, and employed the stellar model atmospheres by \citet{cas03}
for temperature-color transformations.
Readers are referred to \citet{joo13} and references therein for details of the model construction.
In Figure~\ref{fig12}, our synthetic model is compared with the observed CMDs for $r < 30 \,'$.

In this model, we first adopted the metallicity, reddening, and distance modulus from the RR Lyrae stars,
under the assumptions of [$\alpha$/Fe] = 0.3 and the standard helium-enrichment parameter ($\Delta Y / \Delta Z = 2.0$, $Y$ = 0.23 + $Z(\Delta Y / \Delta Z$)).
The age value was then adjusted until the best match between the model and the observed CMD was obtained,
while those parameters from RR Lyrae stars were mostly fixed.
The input parameters used in our best simulation are listed in Table~\ref{tbl3}.
Figure~\ref{fig12} shows that there is no clear sign of young stellar populations,
in addition to the old stellar population reproduced by our model in panel~(b),
indicating that Crater~II is dominated by an old stellar population.
The age obtained ($\sim$10.5\,Gyr) is somewhat younger than our previous estimation of the old populations in the MW GCs \citep{joo13,yoo08}
and the determination of UFDs by \citet{bro12,bro14}.
This is easily understood by the red HB morphology of Crater~II despite its low metallicity.

A crucial question regarding present-day dwarf galaxies is whether they are surviving counterparts of the building blocks that merged to form larger galaxies,
as predicted by the $\Lambda$CDM hierarchical clustering paradigm.
Recent analyses based on the RR Lyrae properties suggest that galaxies like most dwarf companions of the MW are probably not the major contributor to the Galactic halo,
except the most massive ones.
For instance, \citet{zin14} proposed that the preponderance of Oo~I variables in the MW halo, shown in the period-amplitude diagram,
may be explained by accretion of galaxies resembling the massive MW satellites such as LMC, SMC, Fornax, and Sgr,
while minority Oo-int and Oo-II stars may partially come from the systems like low mass dSphs or UFDs.
\citet{ste14} and \citet{fio15} found that, contrary to the MW halo and GCs, there is a complete lack of high-amplitude short-period (HASP) RRab stars
in most dwarf galaxies.
\citet{fio15} have further shown that those stars are only present in the metal-rich systems ([Fe/H]~$\gtrsim -1.5$),
and concluded that only massive dwarf galaxies with a broad metallicity distribution like LMC and Sgr may be primary building block candidates of the MW halo,
with limited contribution from less massive systems like present-day dSphs and UFDs.

In this context, with a lack of the HASP RRab stars (as shown in Figures~\ref{fig9} and \ref{fig10}) and relatively low stellar mass,
Crater~II also may not be directly related to the MW building blocks.
Given its size and surface brightness, however, the presence of Crater~II itself implies that there might be even larger galaxies yet to be discovered,
under the current surface brightness detection limit, as possible massive (and therefore metal-rich) remnants of the building blocks.
Further studies of dwarf galaxies including a more complete census and variable star observations will be important
to better understand the role of dwarf galaxies in the assembly of present-day massive galaxies.

\acknowledgments


We thank the anonymous referee for a number of helpful suggestions.
This research has made use of the KMTNet system operated by the Korea Astronomy and Space Science Institute (KASI)
and the data were obtained at one of three host sites, CTIO in Chile.
H.Jeong acknowledges support from the Basic Science Research Program through the National Research Foundation (NRF) of Korea,
funded by the Ministry of Education (NRF-2013R1A6A3A04064993).
S.C.R was partially supported by the Basic Science Research Program through the NRF of Korea funded by the Ministry of Education (2018R1A2B2006445).
Support for this work was also provided by the NRF to the Center for Galaxy Evolution Research (2017R1A5A1070354).
H.Jerjen acknowledges the support of the Australian Research Council through Discovery Project DP150100862.
This research was made possible through the use of the APASS, funded by the Robert Martin Ayers Sciences Fund.

\clearpage

\clearpage
\begin{deluxetable*}{lcccccccccccc}
\tabletypesize{\scriptsize}
\tablecaption{Pulsation Properties of RR Lyrae Stars in Crater~II\label{tbl1}}
\vspace{-8mm}
\tablehead{
\colhead{ID} & \colhead{R.A.} & \colhead{Dec.} & \colhead{Type} & \colhead{Period} & \colhead{Epoch (max)} & \colhead{$\langle B \rangle$\tablenotemark{a}}
& \colhead{$\langle V \rangle$\tablenotemark{a}} & \colhead{$N_B$\tablenotemark{b}} & \colhead{$N_V$\tablenotemark{b}}
& \colhead{$A_B$\tablenotemark{c}} & \colhead{$A_V$\tablenotemark{c}} & \colhead{Note} \\
\colhead{} & \colhead{(2000)} & \colhead{(2000)} & \colhead{} & \colhead{(days)} & \colhead{($-$2457000)} & \colhead{(mag)} & \colhead{(mag)}
 & \colhead{} & \colhead{} & \colhead{(mag)} & \colhead{(mag)} & \colhead{}
}
\startdata
 V1  &  11:48:46.95  &  $-$18:37:28.97  &  ab  &  0.7633  &  525.8221  &  20.921  &  20.565  &  141  &  144  &  1.155  &  0.807  &  field?\tablenotemark{d} \\
 V2  &  11:48:59.49  &  $-$18:10:09.97  &  ab  &  0.6053  &  762.3792  &  21.360  &  20.999  &  136  &  142  &  1.000  &  0.697  &   \\
 V3  &  11:48:38.76  &  $-$18:31:40.36  &  ab  &  0.6009  &  528.5124  &  21.269  &  20.925  &  138  &  144  &  1.128  &  0.791  &   \\
 V4  &  11:48:51.17  &  $-$18:16:34.26  &   c  &  0.4189  &  765.8670  &  21.289  &  20.945  &  142  &  145  &  0.418  &  0.284  &   \\
 V5  &  11:48:51.88  &  $-$18:31:56.98  &  ab  &  0.5989  &  526.3172  &  21.277  &  20.926  &  141  &  144  &  1.205  &  0.836  &   \\
 V6  &  11:49:04.42  &  $-$18:02:02.16  &   c  &  0.4152  &  759.0594  &  21.224  &  20.897  &  138  &  141  &  0.476  &  0.370  &   \\
 V7  &  11:48:30.82  &  $-$17:41:58.38  &  ab  &  0.5922  &  422.8492  &  21.088  &  20.782  &   54  &   54  &  1.251  &  0.856  &   \\
 V8  &  11:48:47.00  &  $-$18:31:30.74  &  ab  &  0.6496  &  569.1721  &  21.288  &  20.944  &  139  &  144  &  0.746  &  0.502  &   \\
 V9  &  11:46:00.64  &  $-$18:49:38.18  &  ab  &  0.5878  &  421.0890  &  21.257  &  20.937  &   51  &   54  &  0.860  &  0.637  &   \\
V10  &  11:47:51.72  &  $-$18:11:12.19  &  ab  &  0.6212  &  704.6499  &  21.215  &  20.836  &  140  &  144  &  1.000  &  0.721  &   \\
V11  &  11:48:47.71  &  $-$17:58:10.30  &  ab  &  0.6775  &  598.5794  &  21.319  &  20.888  &  141  &  144  &  0.632  &  0.442  &   \\
V12  &  11:48:41.01  &  $-$18:31:45.59  &  ab  &  0.6415  &  673.1422  &  21.401  &  20.998  &  139  &  144  &  0.658  &  0.489  &   \\
V13  &  11:48:51.60  &  $-$18:41:34.13  &  ab  &  0.6474  &  530.1833  &  21.292  &  20.903  &  139  &  144  &  0.849  &  0.649  &   \\
V14  &  11:48:39.63  &  $-$18:09:24.26  &  ab  &  0.6081  &  437.9900  &  21.315  &  20.931  &  139  &  143  &  0.895  &  0.605  &   \\
V15  &  11:47:57.03  &  $-$18:08:39.69  &  ab  &  0.6378  &  488.0886  &  21.309  &  20.899  &  141  &  141  &  0.852  &  0.618  &   \\
V16  &  11:48:33.63  &  $-$18:04:30.77  &  ab  &  0.6023  &  431.0587  &  21.254  &  20.862  &  140  &  144  &  1.189  &  0.850  &   \\
V17  &  11:47:34.18  &  $-$18:39:23.29  &  ab  &  0.6232  &  648.7332  &  21.391  &  20.957  &  140  &  144  &  0.828  &  0.635  &   \\
V18  &  11:47:33.17  &  $-$18:06:30.50  &  ab  &  0.6293  &  684.8898  &  21.369  &  20.945  &  136  &  143  &  0.733  &  0.534  &   \\
V19  &  11:48:04.16  &  $-$18:04:44.49  &  ab  &  0.6303  &  476.6949  &  21.344  &  20.939  &  137  &  144  &  0.824  &  0.569  &   \\
V20  &  11:48:35.40  &  $-$18:30:54.83  &  ab  &  0.6201  &  523.9000  &  21.325  &  20.959  &  137  &  144  &  1.152  &  0.823  &   \\
V21  &  11:49:05.76  &  $-$18:32:03.12  &  ab  &  0.6410  &  440.5893  &  21.487  &  21.064  &  140  &  143  &  0.425  &  0.348  &   \\
V22  &  11:48:14.91  &  $-$18:15:09.20  &  ab  &  0.6001  &  525.2433  &  21.306  &  20.922  &  137  &  143  &  1.071  &  0.770  &   \\
V23  &  11:48:10.64  &  $-$18:17:42.81  &  ab  &  0.6147  &  677.0265  &  21.357  &  20.947  &  139  &  142  &  0.980  &  0.731  &   \\
V24  &  11:48:42.45  &  $-$19:10:16.40  &  ab  &  0.6237  &  575.8158  &  21.470  &  21.061  &  120  &  126  &  0.917  &  0.708  &   \\
V25  &  11:49:05.84  &  $-$18:40:47.01  &  ab  &  0.6083  &  637.7355  &  21.347  &  20.987  &  134  &  144  &  1.058  &  0.738  &   \\
V26  &  11:51:45.15  &  $-$18:29:39.52  &  ab  &  0.7787  &  493.0874  &  20.958  &  20.540  &  124  &  126  &  0.743  &  0.473  &  field?\tablenotemark{d} \\
V27  &  11:50:26.35  &  $-$17:59:03.83  &  ab  &  0.6465  &  737.8831  &  21.284  &  20.879  &  141  &  143  &  0.851  &  0.564  &   \\
V28  &  11:50:15.89  &  $-$18:09:38.24  &  ab  &  0.6239  &  739.1900  &  21.409  &  20.998  &  139  &  144  &  0.879  &  0.629  &   \\
V29  &  11:49:23.50  &  $-$18:20:08.62  &   c  &  0.4212  &  768.2114  &  21.248  &  20.914  &  141  &  144  &  0.485  &  0.344  &   \\
V30  &  11:52:15.27  &  $-$17:59:36.36  &   c  &  0.4291  &  687.0483  &  21.205  &  20.853  &  123  &  126  &  0.532  &  0.369  &   \\
V31  &  11:51:45.12  &  $-$18:07:32.99  &  ab  &  0.6219  &  592.6235  &  21.344  &  20.920  &  122  &  126  &  0.574  &  0.408  &   \\
V32  &  11:50:28.96  &  $-$18:16:44.18  &  ab  &  0.6109  &  697.0341  &  21.270  &  20.922  &  141  &  144  &  1.287  &  0.941  &   \\
V33  &  11:49:45.66  &  $-$18:32:56.41  &   c  &  0.4174  &  467.0077  &  21.288  &  20.953  &  142  &  144  &  0.555  &  0.413  &   \\
V34  &  11:50:44.94  &  $-$18:35:15.51  &  ab  &  0.7293  &  752.6376  &  21.183  &  20.858  &  141  &  144  &  0.692  &  0.550  &  evolved? \\
V35  &  11:49:56.30  &  $-$18:37:43.16  &  ab  &  0.6309  &  422.9369  &  21.288  &  20.898  &  142  &  145  &  0.836  &  0.555  &   \\
V36  &  11:51:01.26  &  $-$18:32:21.17  &  ab  &  0.6775  &  510.8891  &  21.377  &  20.964  &  141  &  143  &  0.517  &  0.379  &   \\
V37  &  11:53:04.48  &  $-$17:55:31.25  &  ab  &  0.5945  &  707.3147  &  21.058  &  20.759  &  121  &  124  &  1.414  &  1.039  &  field? \\
V38  &  11:50:41.80  &  $-$18:40:20.12  &  ab  &  0.6206  &  648.1834  &  21.475  &  21.083  &  140  &  143  &  0.593  &  0.424  &   \\
V39  &  11:49:28.96  &  $-$18:01:34.50  &  ab  &  0.6398  &  777.5090  &  21.330  &  20.919  &  142  &  144  &  0.715  &  0.455  &   \\
V40  &  11:49:46.30  &  $-$18:41:45.65  &  ab  &  0.6288  &  527.1416  &  21.425  &  21.011  &  137  &  143  &  0.995  &  0.714  &   \\
V41  &  11:50:03.59  &  $-$18:44:29.89  &  ab  &  0.6190  &  503.8013  &  21.355  &  20.954  &  138  &  143  &  0.937  &  0.638  &   \\
V42  &  11:49:60.00  &  $-$18:05:32.04  &  ab  &  0.6269  &  508.8869  &  21.466  &  21.047  &  141  &  144  &  0.489  &  0.328  &   \\
V43  &  11:50:41.04  &  $-$18:16:30.60  &  ab  &  0.6245  &  602.3398  &  21.304  &  20.921  &  140  &  144  &  0.925  &  0.661  &   \\
V44  &  11:49:50.68  &  $-$18:45:00.70  &  ab  &  0.6555  &  652.6502  &  21.464  &  21.023  &  141  &  144  &  0.817  &  0.559  &   \\
V45  &  11:50:36.21  &  $-$17:57:29.45  &  ab  &  0.6153  &  500.0881  &  21.271  &  20.883  &  140  &  144  &  1.071  &  0.737  &   \\
V46  &  11:49:36.82  &  $-$18:35:26.76  &  ab  &  0.6155  &  733.8168  &  21.343  &  20.980  &  141  &  144  &  1.006  &  0.739  &   \\
V47  &  11:50:28.10  &  $-$17:52:38.65  &  ab  &  0.6199  &  744.8254  &  21.380  &  20.991  &   97  &   99  &  0.629  &  0.413  &   \\
V48  &  11:50:29.54  &  $-$18:43:02.85  &  ab  &  0.6195  &  654.8936  &  21.362  &  20.991  &  135  &  144  &  1.045  &  0.720  &   \\
V49  &  11:50:07.96  &  $-$18:41:46.42  &  ab  &  0.6157  &  696.5972  &  21.382  &  20.997  &  140  &  143  &  0.932  &  0.647  &   \\
V50  &  11:49:46.73  &  $-$18:32:13.11  &  ab  &  0.5877  &  674.4818  &  21.228  &  20.902  &  140  &  145  &  1.440  &  1.055  &   \\
V51  &  11:50:13.18  &  $-$18:59:39.05  &  ab  &  0.6088  &  708.1373  &  21.327  &  20.981  &   94  &   99  &  0.974  &  0.689  &   \\
V52  &  11:49:40.59  &  $-$18:19:04.64  &  ab  &  0.6280  &  467.7285  &  21.337  &  20.963  &  139  &  142  &  1.133  &  0.801  &   \\
V53  &  11:50:25.11  &  $-$18:44:06.21  &  ab  &  0.5717  &  562.2589  &  21.295  &  20.955  &  138  &  144  &  1.264  &  0.963  &   \\
V54  &  11:52:17.11  &  $-$18:34:52.91  &  ab  &  0.5828  &  558.2980  &  21.289  &  20.957  &  123  &  126  &  1.235  &  0.887  &   \\
V55  &  11:51:48.10  &  $-$18:37:45.56  &  ab  &  0.5524  &  420.8955  &  21.275  &  20.957  &   61  &   62  &  1.210  &  0.929  &   \\
V56  &  11:49:44.86  &  $-$18:35:02.55  &  ab  &  0.5658  &  622.0854  &  21.266  &  20.956  &  139  &  144  &  1.329  &  1.023  &   \\
V57  &  11:49:53.69  &  $-$18:16:39.62  &  ab  &  0.6314  &  466.5745  &  21.465  &  21.077  &  139  &  145  &  0.765  &  0.544  &   \\
V58  &  11:50:06.34  &  $-$18:45:52.32  &  ab  &  0.5739  &  519.9465  &  21.418  &  21.050  &  136  &  143  &  1.148  &  0.840  &   \\
V59  &  11:49:14.69  &  $-$18:46:04.99  &   c  &  0.4440  &  716.7583  &  21.193  &  20.870  &  106  &  109  &  0.543  &  0.363  &   \\
V60  &  11:47:09.71  &  $-$18:22:15.93  &  ab  &  0.6184  &  747.3337  &  21.304  &  20.923  &   86  &   90  &  0.838  &  0.575  &   \\
V61  &  11:49:10.58  &  $-$18:01:03.85  &  ab  &  0.6150  &  509.6780  &  21.332  &  20.916  &  105  &  109  &  0.685  &  0.527  &   \\
V62  &  11:50:22.85  &  $-$18:24:50.27  &  ab  &  0.6106  &  648.3371  &  21.303  &  20.928  &  106  &  109  &  0.805  &  0.593  &   \\
V63  &  11:50:30.79  &  $-$18:21:32.67  &  ab  &  0.6318  &  577.9878  &  21.287  &  20.898  &  104  &  108  &  0.829  &  0.616  &   \\
V64  &  11:50:18.67  &  $-$18:24:41.02  &   c  &  0.4357  &  623.2801  &  21.246  &  20.913  &  105  &  107  &  0.516  &  0.414  &   \\
V65  &  11:47:45.05  &  $-$18:23:26.99  &  ab  &  0.6513  &  426.8200  &  21.343  &  20.909  &  104  &  109  &  0.930  &  0.610  &   \\
V66  &  11:49:47.00  &  $-$18:27:20.97  &  ab  &  0.6268  &  691.0083  &  21.390  &  20.990  &  105  &  108  &  1.057  &  0.740  &   \\
V67  &  11:49:08.33  &  $-$18:48:13.60  &  ab  &  0.5793  &  508.5757  &  21.323  &  20.989  &  103  &  107  &  1.480  &  1.058  &   \\
V68  &  11:49:12.51  &  $-$18:19:36.59  &  ab  &  0.6478  &  584.6497  &  21.424  &  20.989  &  105  &  104  &  0.601  &  0.404  &   \\
V69  &  11:50:42.38  &  $-$18:45:42.65  &   c  &  0.4199  &  748.9645  &  21.411  &  20.993  &  142  &  144  &  0.378  &  0.254  &  less certain \\
V70  &  11:49:12.04  &  $-$18:15:50.36  &  ab  &  0.6245  &  734.3474  &  21.388  &  20.960  &  104  &  108  &  0.924  &  0.647  &   \\
V71  &  11:50:15.91  &  $-$18:25:53.60  &  ab  &  0.6030  &  566.5885  &  21.336  &  20.972  &  104  &  108  &  1.093  &  0.750  &   \\
V72  &  11:48:49.99  &  $-$18:28:27.59  &  ab  &  0.6553  &  590.3775  &  21.428  &  21.004  &  138  &  143  &  0.433  &  0.341  &   \\
V73  &  11:49:30.22  &  $-$18:41:32.33  &  ab  &  0.6493  &  425.8456  &  21.414  &  20.999  &  141  &  143  &  0.754  &  0.539  &   \\
V74  &  11:50:03.94  &  $-$18:26:10.76  &  ab  &  0.6354  &  480.2623  &  21.460  &  21.013  &  105  &  108  &  0.720  &  0.503  &   \\
V75  &  11:50:06.00  &  $-$18:24:33.33  &  ab  &  0.6066  &  455.4853  &  21.271  &  20.917  &  105  &  108  &  1.057  &  0.688  &   \\
V76  &  11:49:04.08  &  $-$18:09:07.73  &  ab  &  0.6516  &  605.3905  &  21.479  &  21.064  &  141  &  145  &  0.527  &  0.357  &   \\
V77  &  11:49:17.38  &  $-$18:38:55.86  &  ab  &  0.6132  &  609.9599  &  21.313  &  20.943  &  104  &  109  &  1.043  &  0.752  &   \\
V78  &  11:49:16.91  &  $-$18:33:54.38  &  ab  &  0.6029  &  494.6237  &  21.382  &  21.015  &  105  &  109  &  0.996  &  0.688  &   \\
V79  &  11:54:43.85  &  $-$17:54:41.34  &  ab  &  0.6727  &  497.5064  &  21.146  &  20.737  &   87  &   90  &  0.861  &  0.600  &  field? \\
V80  &  11:51:51.56  &  $-$18:30:52.50  &   c  &  0.3131  &  452.9847  &  21.186  &  20.818  &  123  &  126  &  0.388  &  0.260  &  less certain \\
V81  &  11:47:08.25  &  $-$18:52:29.86  &  ab  &  0.7251  &  428.6111  &  21.293  &  20.856  &   61  &   62  &  0.619  &  0.376  &  evolved? \\
V82  &  11:47:09.17  &  $-$18:32:22.45  &  ab  &  0.7201  &  771.3640  &  21.140  &  20.829  &  121  &  126  &  0.655  &  0.525  &  evolved? \\
V83  &  11:48:45.47  &  $-$18:09:58.75  &  ab  &  0.6229  &  484.0740  &  21.369  &  20.972  &  139  &  143  &  0.896  &  0.616  &   \\
V84  &  11:46:43.04  &  $-$17:48:50.38  &  ab  &  0.6200  &  421.7759  &  21.308  &  20.948  &   36  &   36  &  0.761  &  0.607  &   \\
V85  &  11:46:02.28  &  $-$18:32:12.10  &  ab  &  0.6065  &  425.7883  &  21.216  &  20.884  &   52  &   54  &  1.270  &  0.800  &   \\
V86  &  11:50:57.17  &  $-$18:46:03.17  &   c  &  0.4116  &  478.0126  &  20.994  &  20.745  &  142  &  145  &  0.814  &  0.590  &   \\
V87  &  11:51:07.05  &  $-$18:39:26.04  &  ab  &  0.6567  &  451.0808  &  21.431  &  21.021  &  140  &  144  &  0.502  &  0.427  &   \\
V88  &  11:49:45.80  &  $-$17:38:45.38  &  ab  &  0.6293  &  425.1559  &  21.213  &  20.828  &   54  &   54  &  1.113  &  0.775  &   \\
V89  &  11:50:49.88  &  $-$17:41:56.21  &  ab  &  0.6262  &  426.0486  &  21.309  &  20.929  &   54  &   54  &  0.560  &  0.440  &   \\
V90  &  11:52:06.09  &  $-$18:44:09.16  &  ab  &  0.7170  &  498.2968  &  21.235  &  20.915  &  122  &  126  &  0.819  &  0.575  &  evolved? \\
V91  &  11:51:28.16  &  $-$18:44:26.91  &   c  &  0.4144  &  425.4381  &  21.318  &  20.982  &   50  &   51  &  0.690  &  0.491  &   \\
V92  &  11:48:58.04  &  $-$18:22:07.78  &  ab  &  0.7084  &  782.6699  &  21.297  &  20.858  &  104  &  109  &  0.473  &  0.317  &   \\
V93  &  11:48:24.47  &  $-$18:23:10.60  &   c  &  0.4056  &  524.7682  &  21.405  &  20.962  &  104  &  109  &  0.327  &  0.252  &  less certain \\
V94  &  11:49:17.63  &  $-$18:06:02.86  &   c  &  0.3970  &  556.2699  &  21.410  &  20.962  &  105  &  109  &  0.255  &  0.200  &  less certain \\
V95  &  11:50:58.40  &  $-$18:27:02.71  &  ab  &  0.6111  &  538.9536  &  21.395  &  20.993  &  105  &  106  &  0.687  &  0.524  &  blend? \\
V96  &  11:52:54.07  &  $-$19:28:31.18  &  ab  &  0.6711  &  759.5477  &  21.432  &  20.974  &   78  &   81  &  0.866  &  0.601  &  field? \\
V97  &  11:49:16.15  &  $-$18:18:35.54  &  ab? &  0.2347  &  439.4544  &  21.636  &  21.334  &  104  &  105  &  0.681  &  0.460  &  field DC? \\
\enddata
\tablenotetext{\rm a}{Intensity-weighted mean magnitude.}
\tablenotetext{\rm b}{Number of observations used in our light curve analysis.}
\tablenotetext{\rm c}{Pulsation amplitude.}
\tablenotetext{\rm d}{field? highly evolved? AC?}
\end{deluxetable*}


\addtolength{\tabcolsep}{15pt}
\begin{deluxetable*}{llc}
\tablecaption{Metallicity, Reddening, and Distance of Crater~II Derived from the RRab Stars\label{tbl2}}
\tablehead{
\colhead{Properties~~~~~~~~~~} & \colhead{Values~~~~~~~}
}
\startdata
$Metallicity,\,\langle \rm[Fe/H] \rangle$  &  $-1.65 \,\pm\, 0.15 \,(0.02)\tablenotemark{a}$              \\
$Reddening,\,\langle E(B-V) \rangle$       &  \phs  $0.05 \pm\, 0.02 \,(0.003)\tablenotemark{a}$          \\
$Distance~Modulus,\,(m-M)_0$ (mag)         &  \phd $20.25 \,\pm\, 0.10$                                   \\
$Distance,\,\rm d_\odot$ (kpc)             &  \phs \phd $112 \,\pm\, 5$                                   \\
\enddata
\tablenotetext{\rm a}{The uncertainty is the standard deviation and the number in parentheses is the standard error of the mean.}
\end{deluxetable*}
\addtolength{\tabcolsep}{-15pt}

\begin{deluxetable*}{lcccccc}
\tabletypesize{\scriptsize}
\tablecaption{Input Parameters Used in Our Best Simulation of the Crater~II Stellar Population\label{tbl3}}
\tablewidth{0pt}
\tablehead{
\colhead{$Z$} & \colhead{[Fe/H]\tablenotemark{a}} & \colhead{$Y$\tablenotemark{b}}
& \colhead{Age} & \colhead{Mass-loss\tablenotemark{c}} & \colhead{$\sigma _M$\tablenotemark{d}}         \\
\colhead{} & \colhead{} & \colhead{} & \colhead{(Gyr)} & \colhead{($M_\odot$)} & \colhead{($M_\odot$)}
}
\startdata
0.00071 & $-$1.65 & 0.231  & 10.5  & 0.187 & 0.02  \\
\enddata
\tablenotetext{\rm a}{[$\alpha$/Fe] = 0.3.}
\tablenotetext{\rm b}{From the standard helium-enrichment parameter, i.e., $\Delta Y / \Delta Z = 2.0,~ Y = 0.23 + Z(\Delta Y / \Delta Z$).}
\tablenotetext{\rm c}{Mean mass-loss on the RGB for \citet{rei77} mass-loss parameter, $\eta$ = 0.53.}
\tablenotetext{\rm d}{Mass dispersion on the HB.}
\end{deluxetable*}

\begin{figure*}
\epsscale{1.0}
\plotone{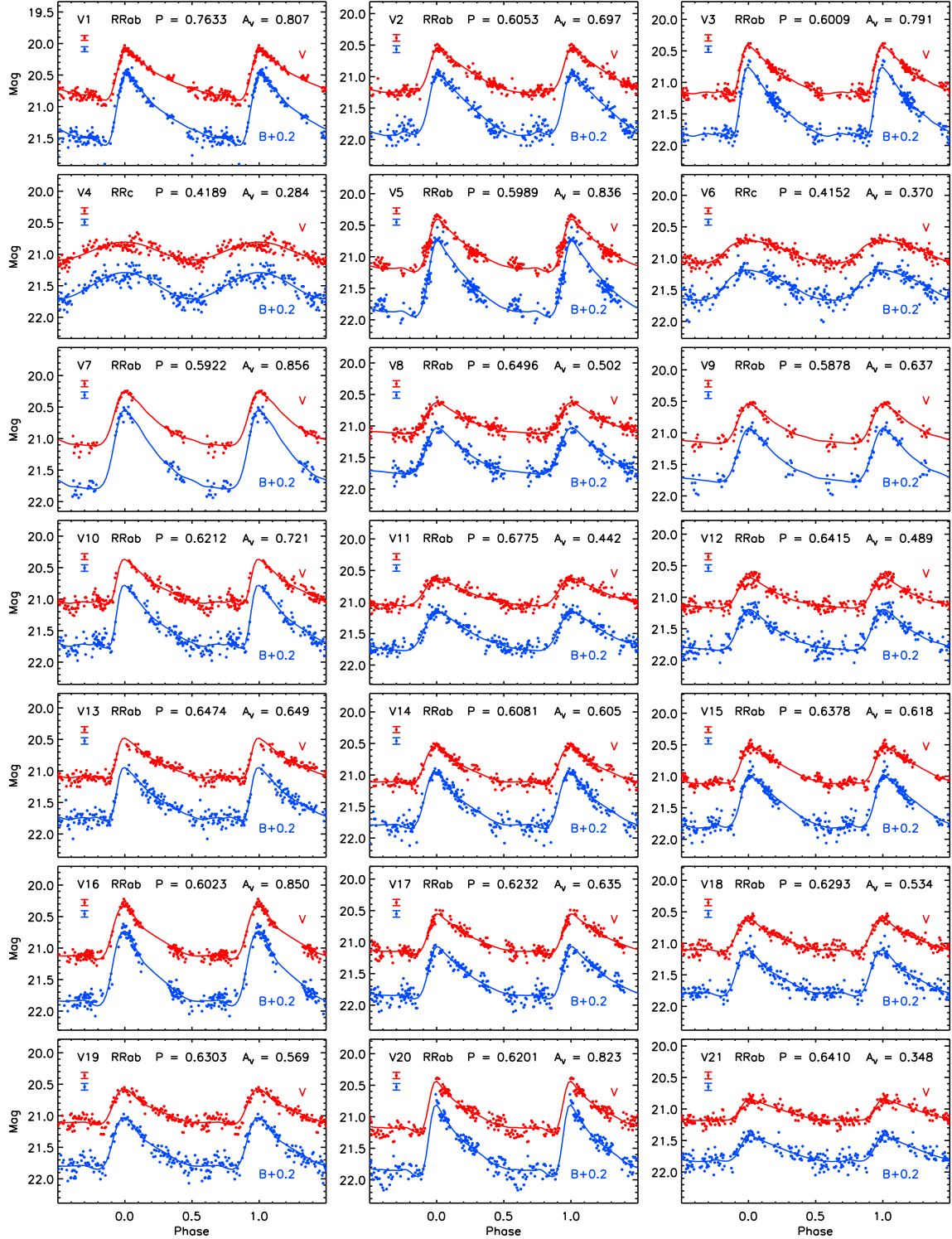}
\caption{
Light curves of RR Lyrae stars for $B$- and $V$-bands (blue and red points), respectively, together with our best-fitting templates (solid lines).
The $B$-band data are shifted by +0.2\,mag to separate from the $V$-band.
The variable ID, type, period (day), and $V$-band amplitude (mag) are denoted in each panel.
The error bars in the upper left corner represent the mean photometric errors for each band.
\label{fig2}}
\end{figure*}

\begin{figure*}
\epsscale{1.0}
\plotone{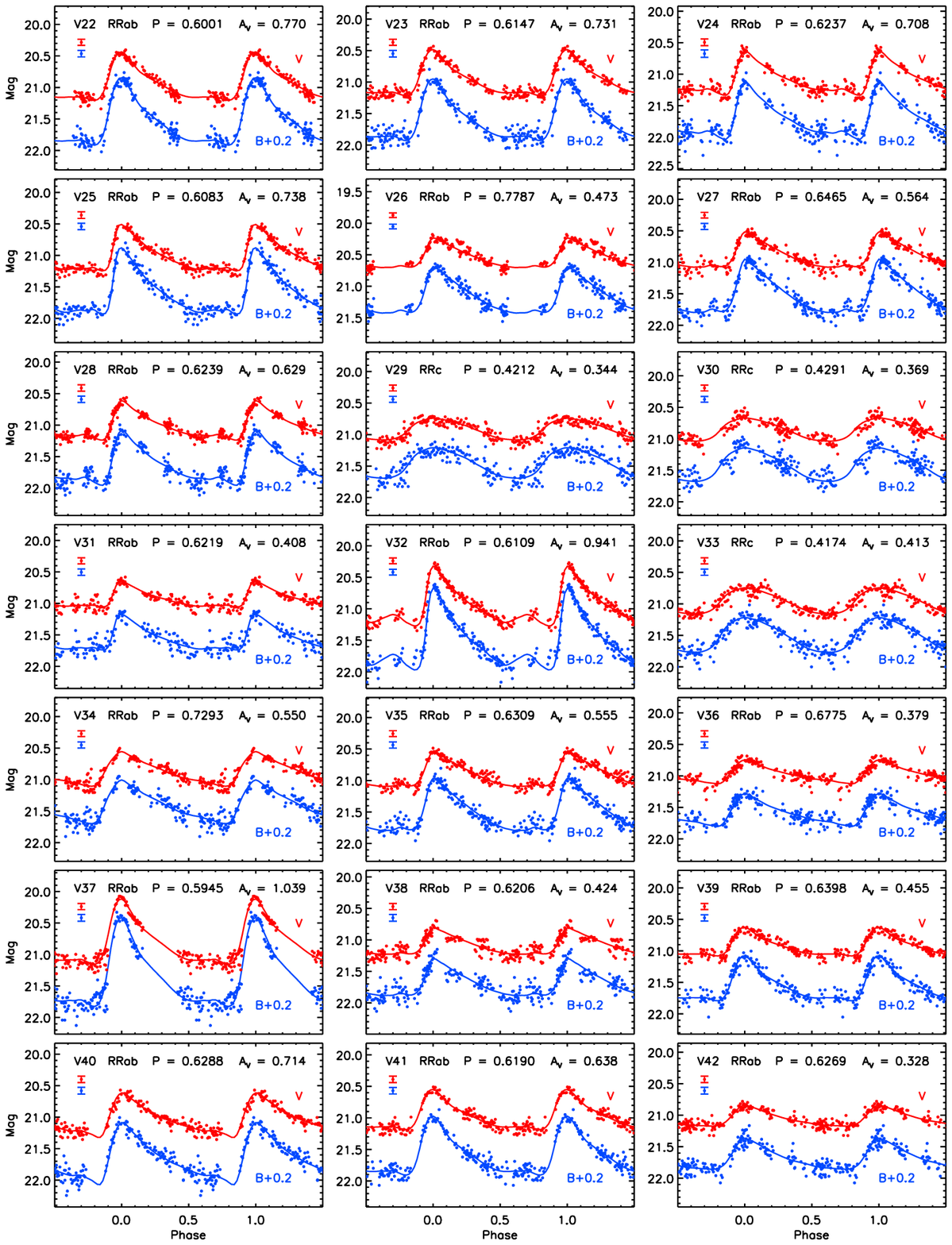}
\caption{
Continued from Figure~\ref{fig2}.
\label{fig3}}
\end{figure*}

\begin{figure*}
\epsscale{1.0}
\plotone{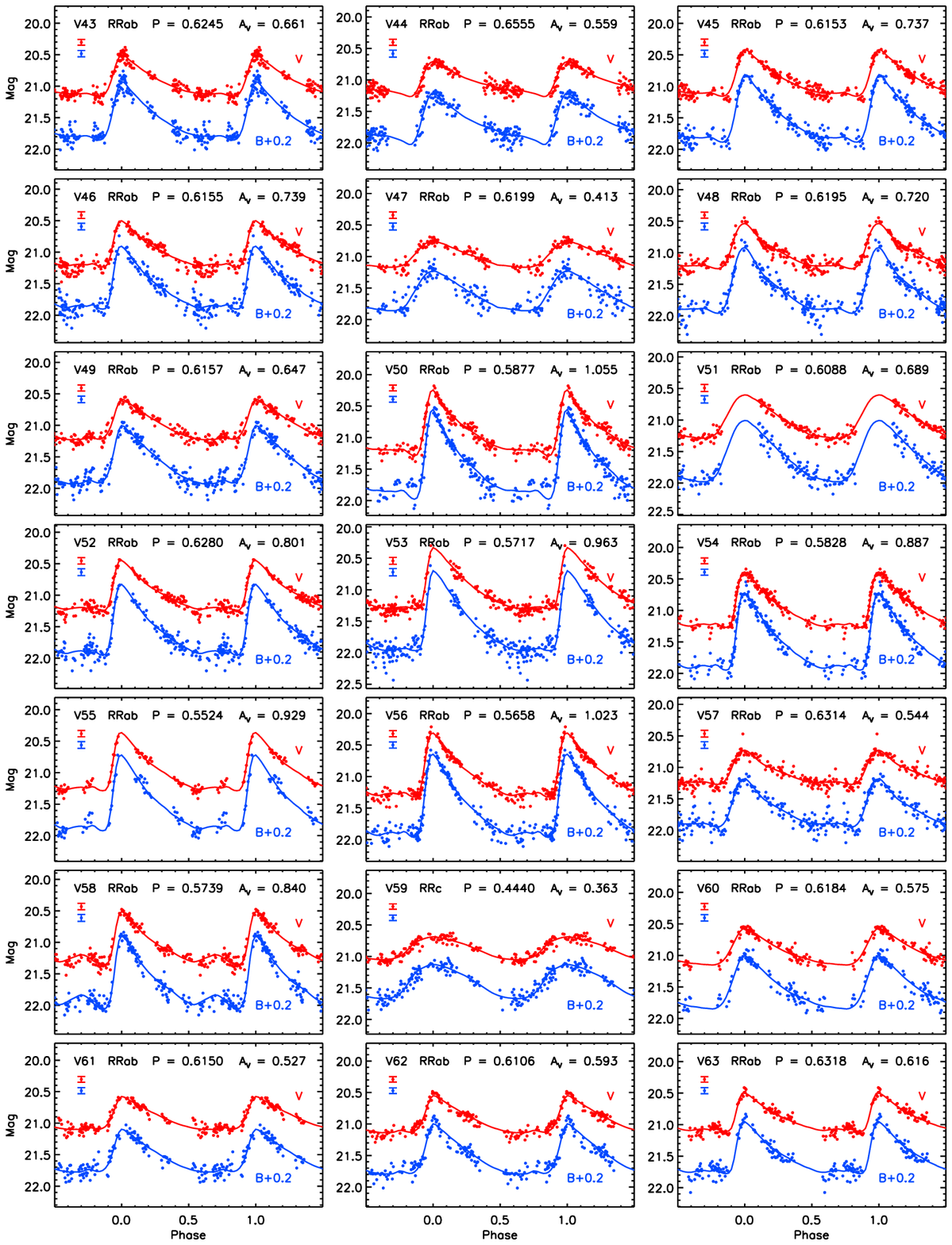}
\caption{
Continued from Figure~\ref{fig3}.
\label{fig4}}
\end{figure*}

\begin{figure*}
\epsscale{1.0}
\plotone{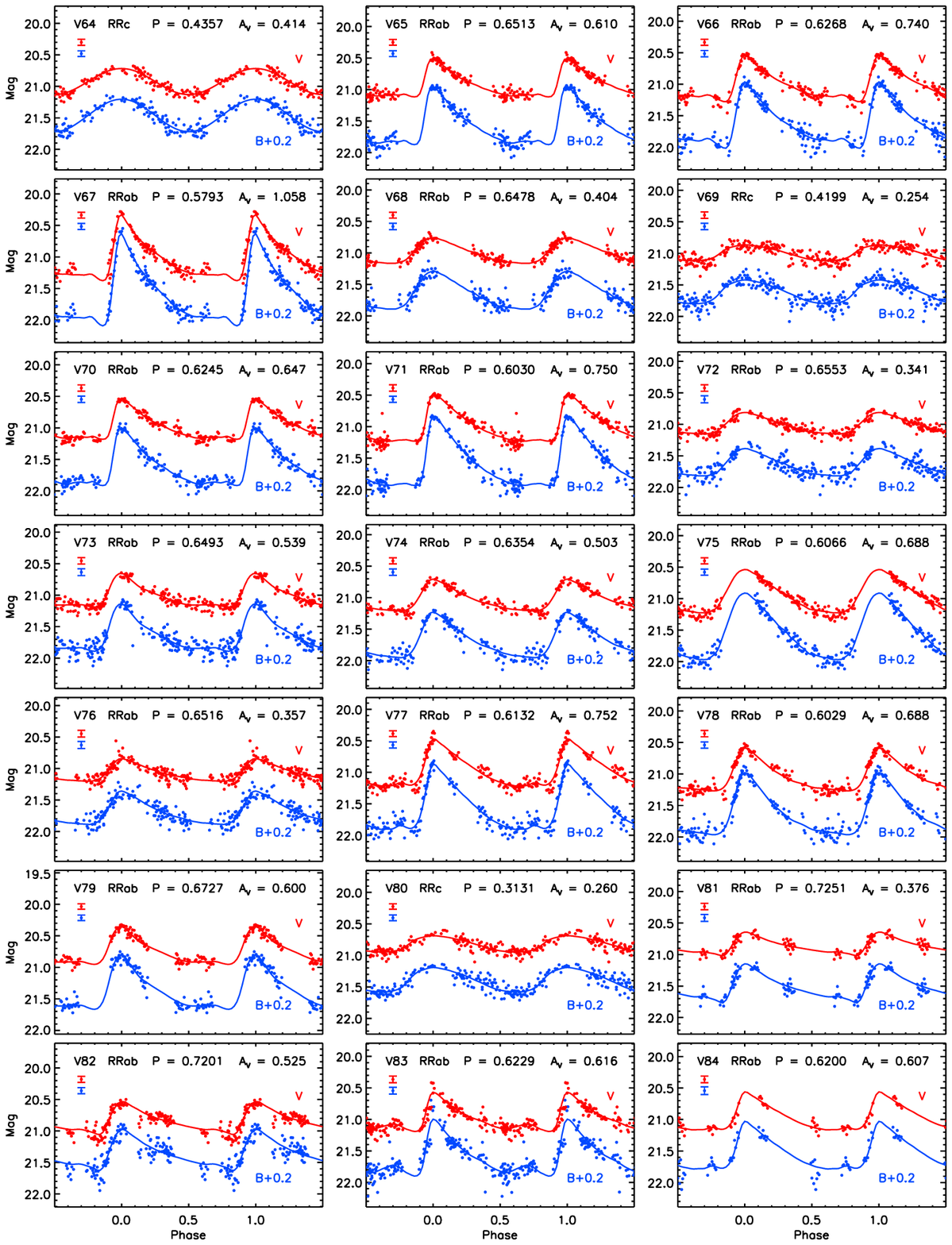}
\caption{
Continued from Figure~\ref{fig4}.
\label{fig5}}
\end{figure*}

\begin{figure*}
\epsscale{1.0}
\plotone{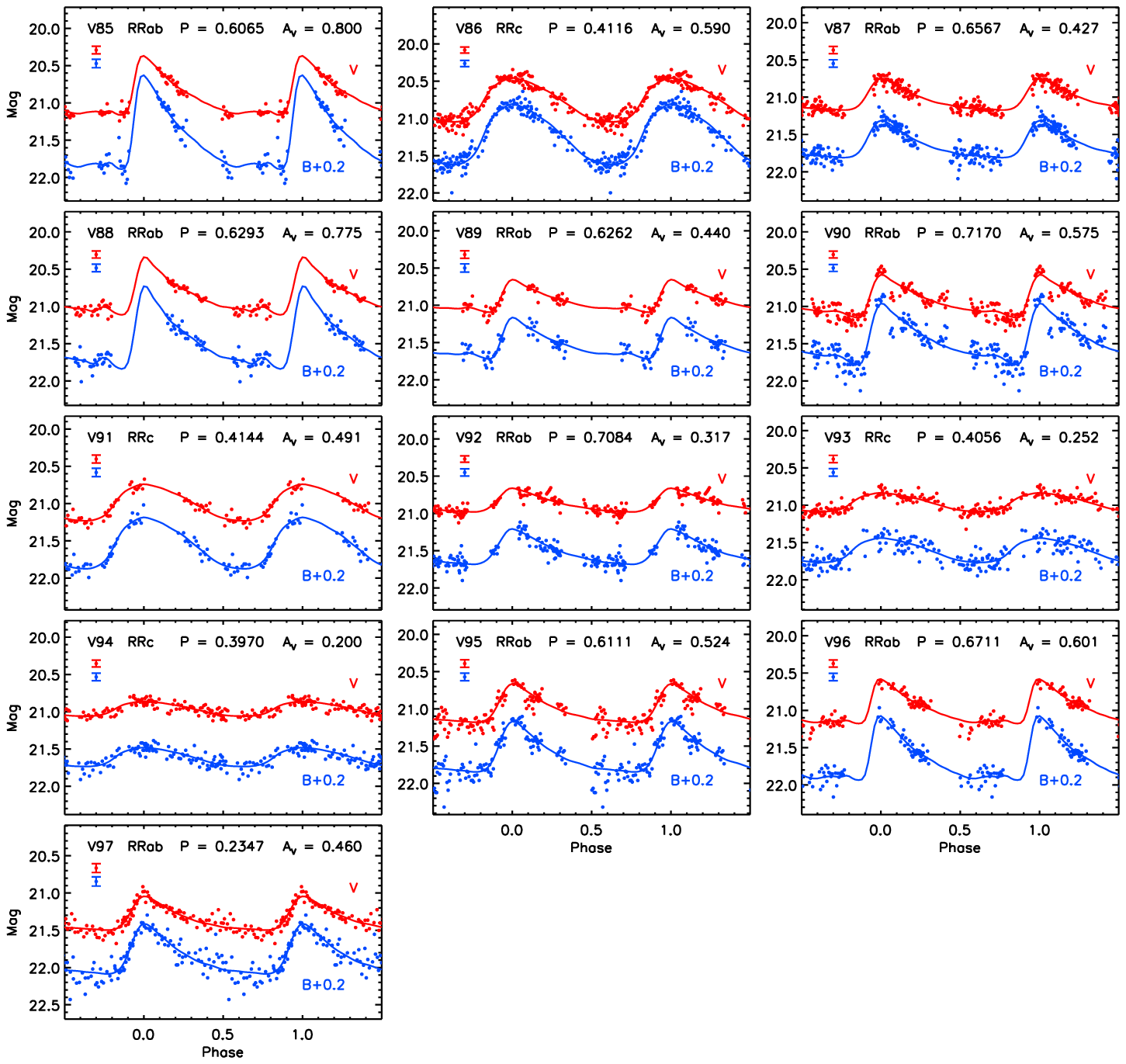}
\caption{
Continued from Figure~\ref{fig5}.
\label{fig6}}
\end{figure*}

\begin{figure*}
\epsscale{0.7}
\plotone{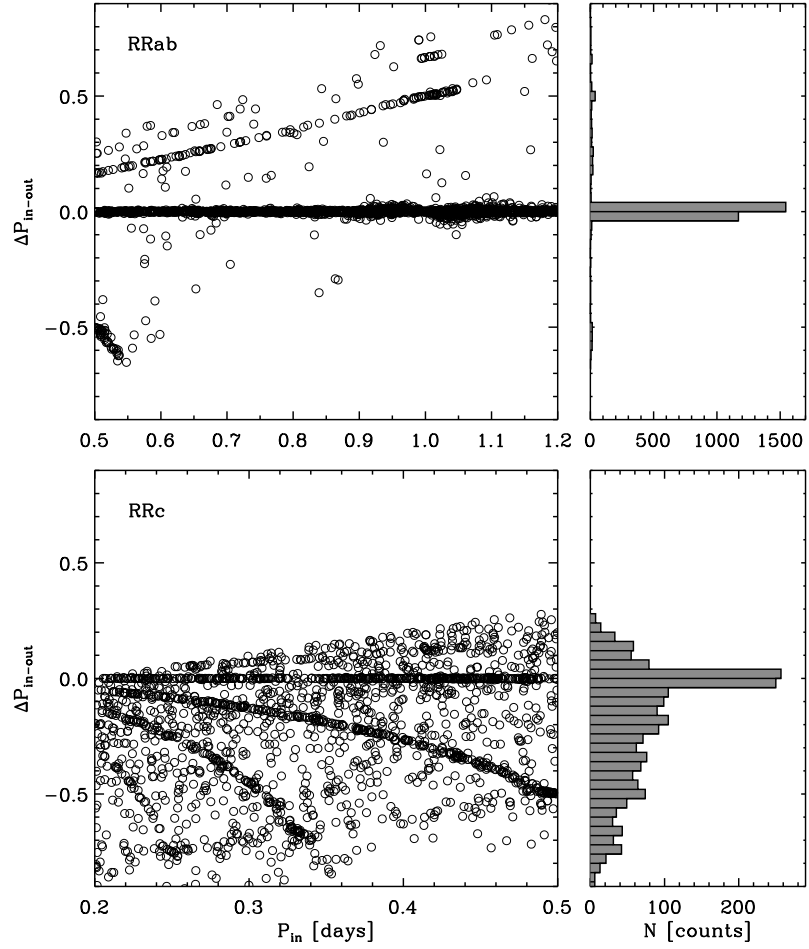}
\caption{Our synthetic light curve simulations for Crater~II, showing the differences between the input and output periods
($\Delta P_{\rm in - out}$) as a function of the input period (left panels) and their distributions (right panels), for the two different pulsation modes.
The simulations indicate that $c\,$-type RR Lyrae stars are much more affected by aliasing compared to $ab\,$-type variables.
\label{fig7}}
\end{figure*}

\begin{figure*}
\epsscale{0.7}
\plotone{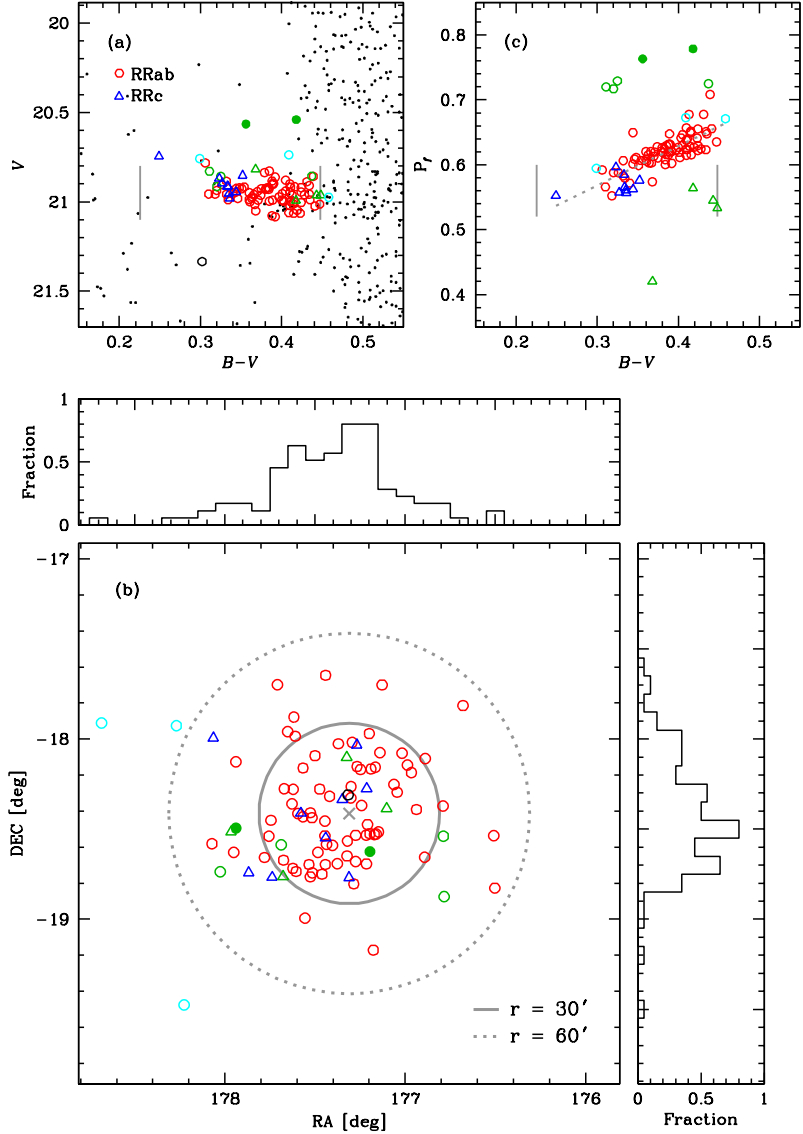}
\caption{
(a) CMD zoomed around the HB region, (b) spatial distribution, and (c) color-period diagram of the RR Lyrae stars, showing that
most of them belong to Crater~II. Circles and triangles denote $ab\,$- and $c\,$-type variables, respectively.
The two vertical grey lines in panels~(a) and (c) represent the empirical instability strip estimated from the nine Galactic and LMC clusters in \citet{wal98},
which are also reddened by 0.05\,mag (see the text).
The three variables (cyan circles) outside $60\,'$ from the center of the galaxy (grey cross) might be field RR Lyrae stars.
The periods of RRc stars in panel~(c) are fundamentalized assuming $P_c / P_{ab}$ = 0.745 \citep{cle01,nem85}.
The dotted line in panel~(c) is the robust linear fit to the data, where 10 RR Lyrae stars are outside the 3\,$\sigma$ range.
The two brightest variables with the longest periods (green filled circles) might be either field RR Lyrae stars, highly evolved RR Lyrae stars, or ACs.
The four RRab stars (green open circles) with long periods appear to be evolved RR Lyrae stars, while the four RRc stars (green triangles)
with short periods are less certain. A probable non-RR Lyrae variable, V97, is marked as a black circle in panels~(a) and (b).
\label{fig8}}
\end{figure*}

\begin{figure*}
\epsscale{0.5}
\plotone{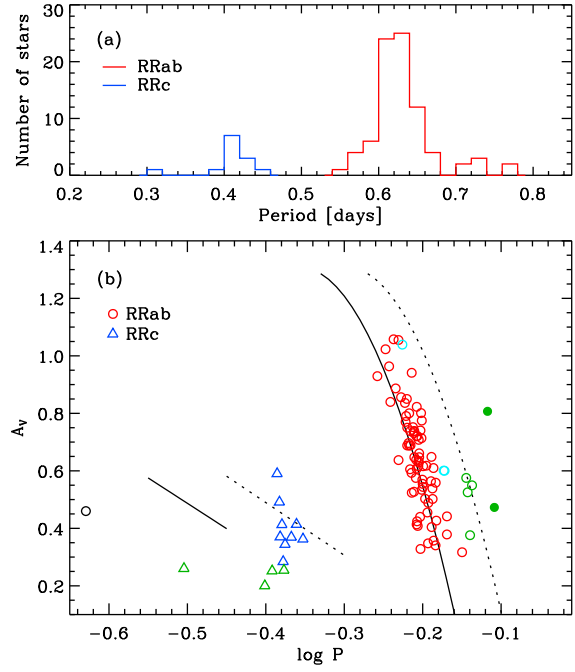}
\caption{
(a) Period distribution and (b) period-amplitude diagram of the RR Lyrae stars in Crater~II.
The solid and dotted lines in panel~(b) represent the loci of the MW Oo~I and Oo~II GCs, respectively \citep{zor10,cac05}.
Symbols are the same as in Figure~\ref{fig8}.
\label{fig9}}
\end{figure*}

\begin{figure*}
\epsscale{1.0}
\plotone{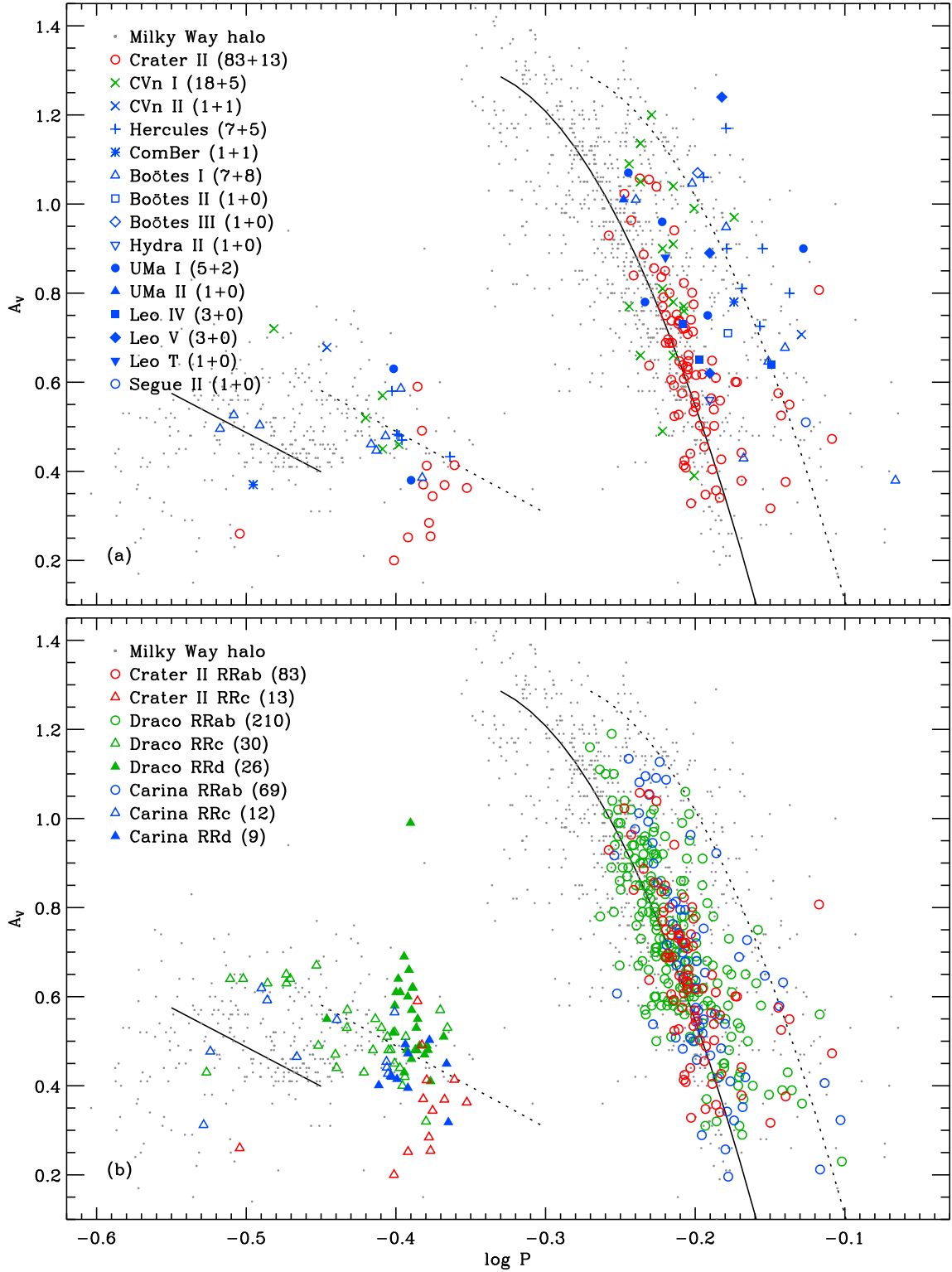}
\caption{
(a) Same as Figure~\ref{fig9}(b), but here the RR Lyrae stars in Crater~II (red circles) are compared with those in the MW halo (grey dots),
CVn~I (green crosses), and the other 13 UFDs (blue symbols). Note that this is just an update of Figure~10 in \citet{viv16}.
Numbers in parentheses are the numbers of $ab\,$- and $c\,$-type variables separated by plus signs.
(b) Same as panel~(a) but compared with those in the two classical dwarfs, Draco (green) and Carina (blue). Numbers in parentheses are the numbers of variables.
\label{fig10}}
\end{figure*}

\begin{figure*}
\epsscale{0.5}
\plotone{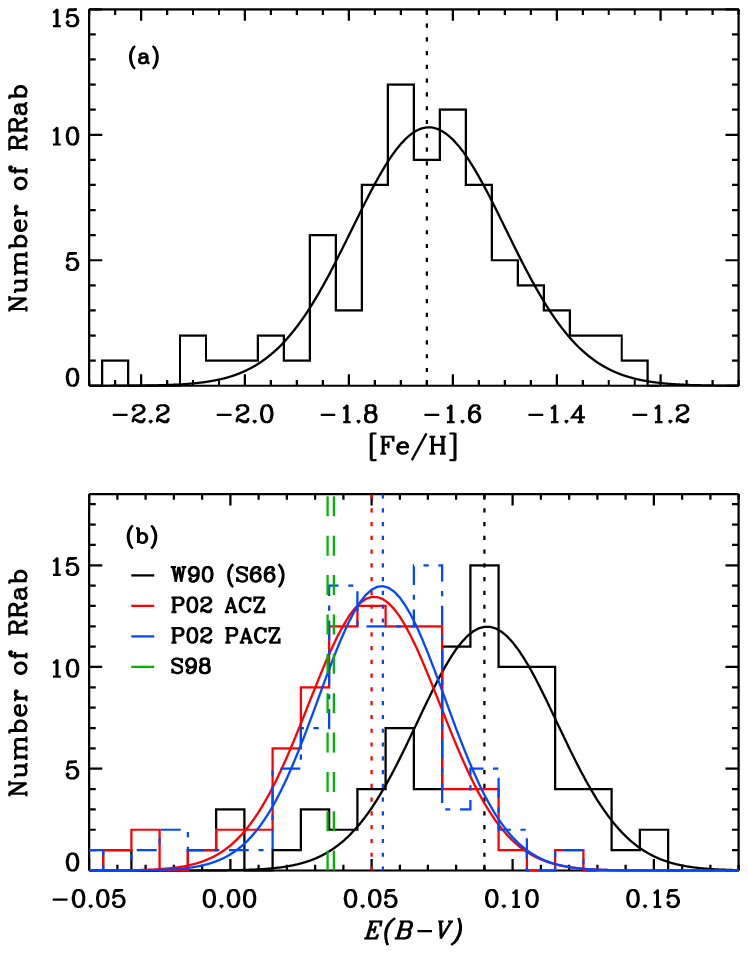}
\caption{
(a) Metallicity distribution of individual RRab stars derived from the equation of \citet{alc00} and a Gaussian fit (solid curve).
(b) Reddening distributions of RRab stars obtained from the equations of \citet[W90, based on \citealt{stu66}, S66]{wal90} and \citet[P02, ACZ and PACZ]{pie02}
and Gaussian fits to the data (solid curves). Note that the Sturch's method tends to overestimate the reddening by $\sim$0.3\,mag (see the text).
The two dashed green lines denote the mean reddenings from the map of \citet[S98]{sch98} for $\rm r < 30\,'$ and $\rm r < 60\,'$ regions around the center of the galaxy.
Dotted lines indicate the peaks of the distributions.
The peak values of metallicity and reddenings hardly change even if the nine $ab\,$-type outliers (i.e., the three stars outside $2\,r_{h}$ and
the six stars with long periods) are excluded.
\label{fig11}}
\end{figure*}

\begin{figure*}
\epsscale{1.0}
\plotone{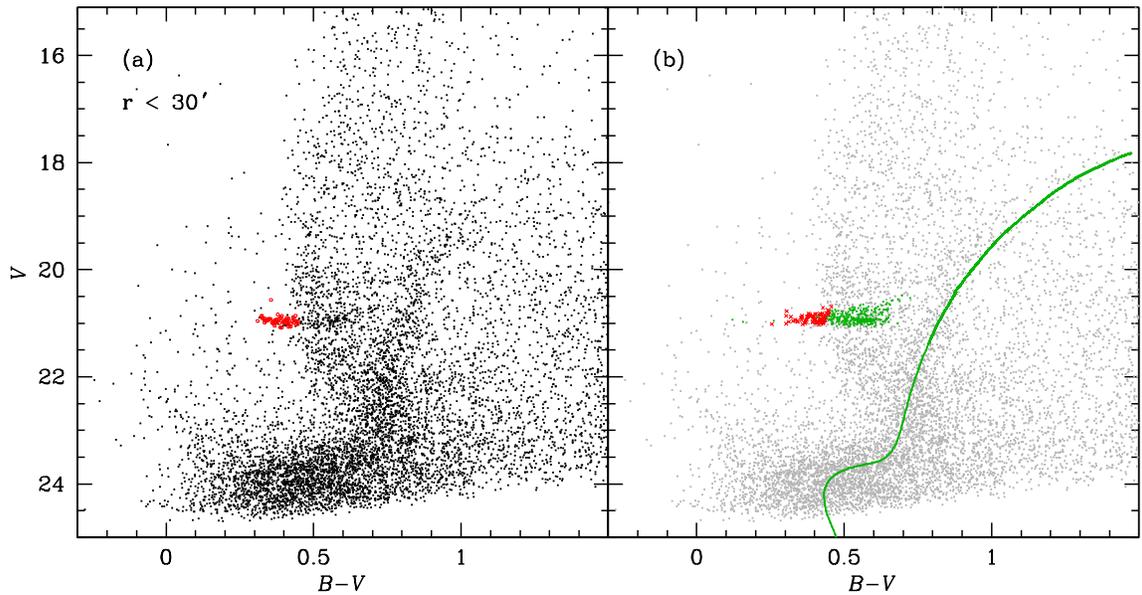}
\caption{
Comparison of our population models with the observations of Crater~II.
(a) Observed CMD, similar to Figure~\ref{fig1}(a) with red open circles for the RR Lyrae stars.
(b) Our population models (green lines and points) on the observed CMD (grey points). Red crosses denote the model RR Lyrae stars.
Parameters used in our simulation are listed in Table~\ref{tbl3}.
Adopted reddening and distance modulus are $E(B-V) = 0.05$ and $(m-M)_0 = 20.30$\,mag, where $(m-M)_0$ was slightly revised by +0.05\,mag (see Table~\ref{tbl2}).
\label{fig12}}
\end{figure*}

\end{document}